\documentclass[oldversion]{aa}
\pdfoutput=1
\usepackage[tbtags,fleqn]{amsmath}
\usepackage{amsfonts}
\usepackage{pstricks, pst-plot}
\usepackage{graphicx}
\usepackage{natbib}

\newcommand{\diff}{\mathrm d}

\newcommand{\mincir}{\raise
  -2.truept\hbox{\rlap{\hbox{$\sim$}}\raise5.truept \hbox{$<$}\ }}
\newcommand{\magcir}{\raise
  -2.truept\hbox{\rlap{\hbox{$\sim$}}\raise5.truept \hbox{$>$}\ }}


\begin{document}

\title{2MASS wide field extinction maps: II. The Ophiuchus and the
  Lupus cloud complexes}
\titlerunning{2MASS extinction maps: II. Ophiuchus and Lupus}
\author{Marco Lombardi\inst{1,2}, Charles J. Lada\inst{3}, and Jo\~ao
  Alves\inst{4}}
\authorrunning{M. Lombardi \textit{et al}.}
\offprints{M. Lombardi}
\mail{mlombard@eso.org}
\institute{%
  European Southern Observatory, Karl-Schwarzschild-Stra\ss e 2, 
  D-85748 Garching bei M\"unchen, Germany 
  \and 
  University of Milan, Department of Physics, via Celoria 16, I-20133
  Milan, Italy (on leave) 
  \and
  Harvard-Smithsonian Center for Astrophysics, Mail Stop 42, 60 Garden
  Street, Cambridge, MA 02138
  \and 
  Calar Alto Observatory -- Centro Astron\'omico Hispano Alem\'an,
  C/Jes\'us Durb\'an Rem\'on 2-2, 04004 Almeria, Spain}
\date{Received ***date***; Acceptet ***date***} 

\abstract{%
  We present an extinction map of a $\sim 1\,700 \mbox{ deg sq}$
  region that encloses the Ophiuchus, the Lupus, and the Pipe dark
  complexes using 42 million stars from the Two Micron All Sky Survey
  (2MASS) point source catalog.  The use of a robust and optimal
  near-infrared method to map dust column density (\textsc{Nicer},
  described in \citealp{2001A&A...377.1023L}) allow us to detect
  extinction as low as $A_K = 0.05 \mbox{ mag}$ with a 2-$\sigma$
  significance, and still to have a resolution of $3 \mbox{ arcmin}$
  on our map.  We also present a novel, statistically sound method to
  characterize the small-scale inhomogeneities in molecular clouds.
  Finally, we investigate the cloud structure function, and show that
  significant deviations from the results predicted by turbulent
  models are observed.

  }

\maketitle

%
%_____________________________________________________________________

\defcitealias{2001A&A...377.1023L}{Paper~I}
\defcitealias{2006A&A...454..781L}{Paper~II}

\section{Introduction}
\label{sec:introduction}

In this paper we present an extinction map of the Ophiuchus and Lupus
complexes covering $\sim 1\,700 \mbox{ deg sq}$, computed by applying
an optimized multi-band technique dubbed Near-Infrared Color Excess
Revisited (\textsc{Nicer} \citealp{2001A&A...377.1023L}, hereafter
Paper~I) to 42 million JHK photometric measurements of stars from the
Two Micron All Sky Survey (2MASS; \citealp{1994ExA.....3...65K}).
This paper is the second of a series where we apply the \textsc{Nicer}
method to point sources from the 2MASS database.  The main aim of this
coordinated study is to investigate in detail the large-scale
structure of molecular clouds.  In addition, the use of a uniform
dataset and of a consistent and well tested pipeline allows us to
characterize many properties of molecular clouds, such as their
reddening law, and to identify cloud-to-cloud variations in such
properties.  The region considered here encloses in addition the Pipe
nebula, considered in the first paper of this series
(\citealp{2006A&A...454..781L}; hereafter Paper~II), and allows us to
study in detail the relationships among these molecular clouds.

Lupus is a well studied complex, composed of several subclouds showing
different modes of star formation: Lupus 1 shows isolated star
formation, while cluster formation is observed in Lupus 3
\citep{1977ApJS...35..161S, 1994AJ....108.1071H, 2000AJ....119..873N};
no evident star formation activity is observed in Lupus 5.  The
distance of this cloud complex is still highly controversial.  The
conventional value, $150 \mbox{ pc}$ \citep{1991ESOSR..11....1R}, has
been challenged by \citet{1998A&A...338..897K}, who reported $100
\mbox{ pc}$, and by \citet{1998MNRAS.301L..39W}, $190 \mbox{ pc}$,
both using Hipparcos observations.  In addition,
\citet{2001A&A...373..714K} suggested that Lupus~2 is physically
disconnected from the other Lupus subclouds, and is located at a
distance of $360 \mbox{ pc}$; similarly, \citet{1999A&A...352..574B}
suggested Lupus~3 to be disconnected from the rest of the Lupus
complex.  However, surprisingly no velocity difference is found among
the various Lupus subclouds in radio observations.  Recently, we used
a novel maximum-likelihood technique based on Hipparcos data obtaining
$d = (155 \pm 8) \mbox{ pc}$ \citep{2008A&A...480..785L}; this value will be
adopted in this paper as the actual distance of Lupus.

Due to its large area on the sky, millimeter observations have mostly
covered limited regions of this cloud.  The first extensive cloud
survey in ${}^{12}$CO ($J=1 \rightarrow 0$) was made by
\citet{1986A&A...167..234M} over $\sim 170 \mbox{ sq deg}$ with an
effective resolution of $30 \mbox{ arcmin}$.  More recently,
\citet{2001PASJ...53.1081T} have performed a complete survey of the
Lupus complex (also in ${}^{12}$CO, $J=1 \rightarrow 0$) using the
NANTEN sub-millimeter telescope.  Their data covered $~ 550 \mbox{ sq
  deg}$ on a grid of $8 \mbox{ arcmin}$, and smaller areas on a $4
\mbox{ arcmin}$ grid.  Finally, recently \citet{2005ApJ...629..276T}
studied the structure of Lupus~3 in the NIR.

The $\rho$ Ophiuchi molecular cloud is a filamentary complex located
at the edge of the Upper Scorpius subgroup in the Sco-Cen OB
association.  It is composed of a series of dark clouds that extend
eastward from cores of dense molecular gas
\citep{1992A&A...262..258D}.  With an estimated distance of $(119 \pm
6) \mbox{ pc}$ \citep{2008A&A...480..785L}, it is one of the nearest
star-forming regions, and has thus been the target of numerous
investigations.  The main cloud, L1688, is situated about one degree
south of the star $\rho$ Ophiuchi and has been studied in the near and
far infrared, in the millimeter continuum, as well as in the X-ray and
radio continuum, and hosts an embedded infrared cluster with $\sim
200$ young stellar objects (YSO; e.g., \citealp{1992ApJ...395..516G,
  2001A&A...372..173B, 1993ApJ...416..185C, 2000PASJ...52.1147T}).
L1688 is known to host a $1 \mbox{ pc} \times 2 \mbox{ pc}$ central
molecular core with visual estimated visual extinction exceeding $50
\mbox { mag}$ \citep{1983ApJ...274..698W}.  This ``main cloud'' of Oph
covers an area of roughly $480 \mbox{ sq arcmin}$ and has been
dissected into about a dozen cloud components.  The entire star
formation complex extends over several degrees on the sky, containing
a few major clouds.  \citet{1998A&A...336..150M} have conducted an
extensive $1.3 \mbox{ mm}$ continuum mapping of the central region
using the IRAM 30-m telescope.  \citet{2006A&A...447..609S} presents
SIMBA $1.2 \mbox{ mm}$ observations of the cloud covering an area of
$4600 \mbox{ arcmin}^2$ with a resolution of $24''$;
\citet{2000ApJ...545..327J} have published a SCUBA $850 \ \mu\mbox{m}$
thermal emission map of the $\rho$ Ophiuchi cloud.

This paper is organized as follows.  In
Sect.~\ref{sec:nicer-absorpt-map} we briefly describe the technique
used to map the dust and we present the main results obtained.  A
statistical analysis of our results and a discussion of the bias
introduced by foreground stars and unresolved substructures is
presented in Sect.~\ref{sec:statistical-analysis}.
Section~\ref{sec:mass-estimate} is devoted to the mass estimate of the
cloud complex.  Finally, we summarize the results obtained in this
paper in Sect.~\ref{sec:conclusions}.

\section{\textsc{Nicer} extinction map}
\label{sec:nicer-absorpt-map}

We carried out the data analysis using the \textsc{Nicer} method
described in Paper~I.  Near infrared $J$ ($1.25\ \mathrm{\mu m}$),
$H$ ($1.65 \ \mathrm{\mu m}$), and $K_\mathrm{s}$ band ($2.17 \
\mathrm{\mu m}$) magnitudes of stars in a large region of the sky
which includes the Ophiuchus and the Lupus dark clouds were taken from
the Two Micron All Sky Survey\footnote{See
  \texttt{http://www.ipac.caltech.edu/2mass/}.}
\citep[2MASS;][]{1994ExA.....3...65K}.  In particular, we selected all
2MASS reliable point sources within the boundaries
\begin{align}
  \label{eq:1}
  -32^\circ <{} & l < +16^\circ \; , & +2^\circ <{} & b < +40^\circ \; .
\end{align}
This area is $1\,672$ square degrees and contains approximately $42$
million point sources from the 2MASS catalog.  Note that since the
region spans several degrees in galactic latitude, the local density
of background stars changes significantly in the field; as a result,
since we used a fixed size for the smoothing of the extinction map
(see below), the noise of the map increases as $b$ increases.

\begin{figure}[!t]
  \centering
  \includegraphics[width=\hsize,bb=13 16 300 299]{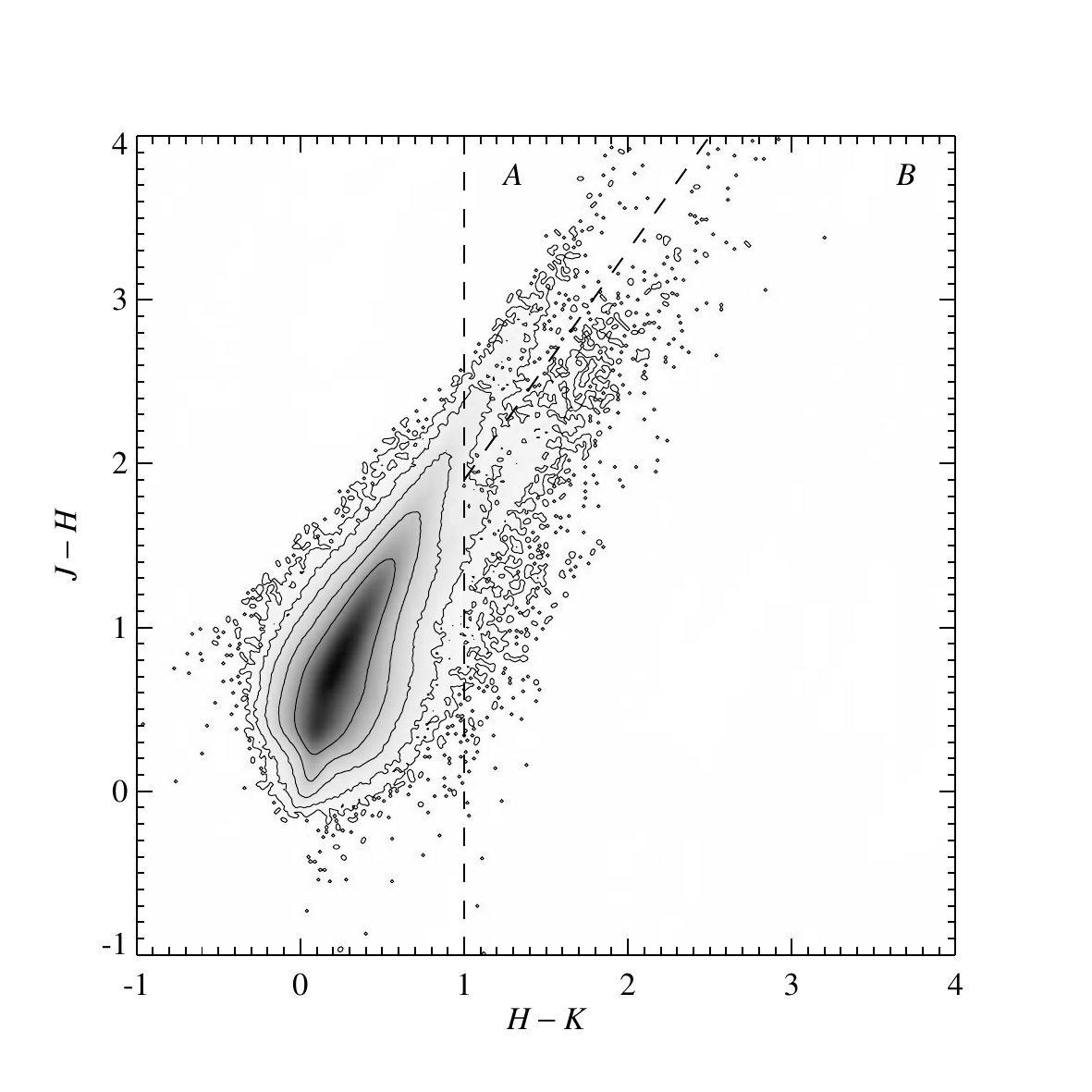}
  \caption{Color-color diagram of the stars in the whole field, as a
    density plot.  The contours are logarithmically spaced, i.e.\ each
    contour represents a density ten times larger than the enclosing
    contour; the outer contour detects single stars and clearly shows
    a bifurcation at large color-excesses.  The dashed lines identify
    the regions in the color space defined in Eqs.~\eqref{eq:2} and
    \eqref{eq:3}, as indicated by the corresponding letters.  Only
    stars with accurate photometry in all bands (maximum 1-$\sigma$
    errors allowed $0.1 \mbox{ mag}$) have been included in this
    plot.}
  \label{fig:1}
\end{figure}

\begin{figure}[!t]
  \centering
  \includegraphics[width=\hsize]{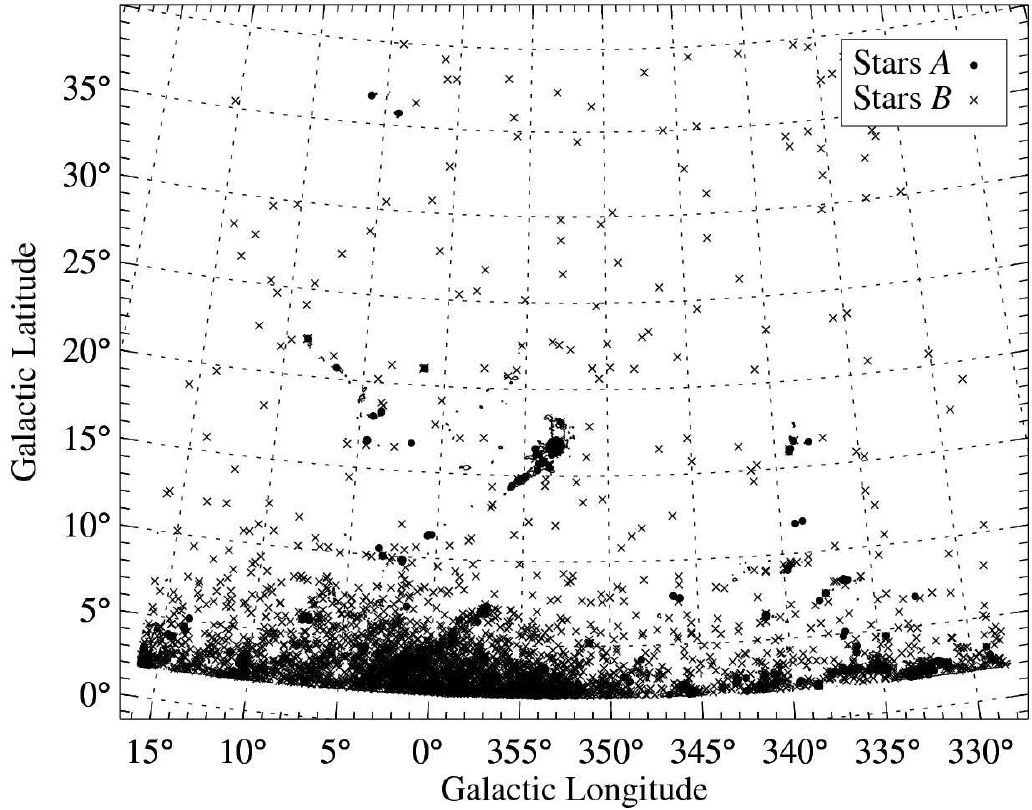}
  \caption{Spatial distribution of the samples of sources as defined
    by Eqs.~\eqref{eq:2} and \eqref{eq:3}.  Sample $A$ is shown as
    filled circles, while sample $B$ is shown as crosses (see also
    Fig.~\ref{fig:1}).  Sample $A$ appears to be strongly clustered in
    high-column density regions of the cloud, and is thus interpreted
    as genuine reddened stars; sample $B$ seems not to be associated
    with the cloud, and is instead preferentially located at low
    galactic latitudes and in the Galactic bulge.}
  \label{fig:2}
\end{figure}

\begin{figure}[!t]
  \centering
  \includegraphics[width=\hsize, bb=17 7 330 240]{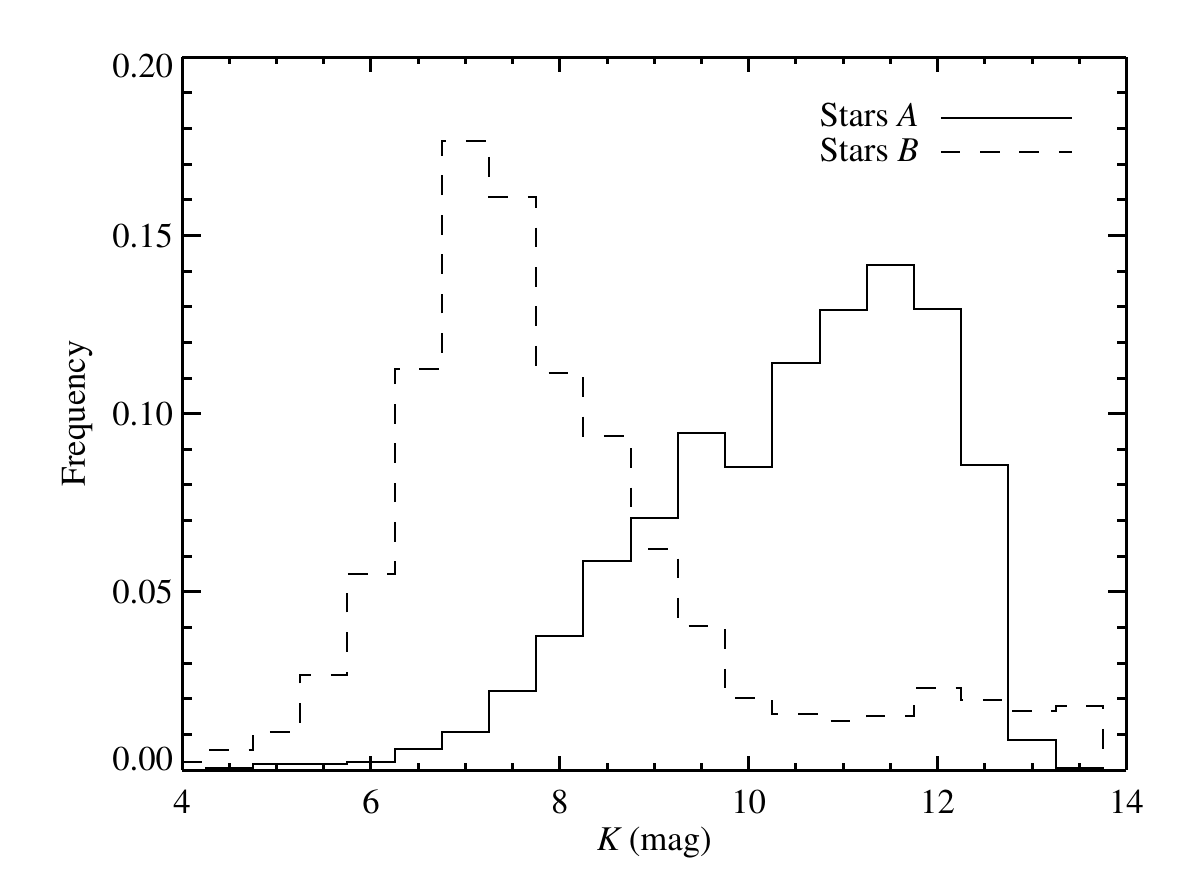}
  \caption{The histogram of the $K$ band magnitude for the two star
    subsets $A$ and $B$ of Eqs.~\eqref{eq:2} and \eqref{eq:3}.}
  \label{fig:3}
\end{figure}

\begin{figure}[!t]
  \centering
  \includegraphics[width=\hsize, bb=13 16 300 299]{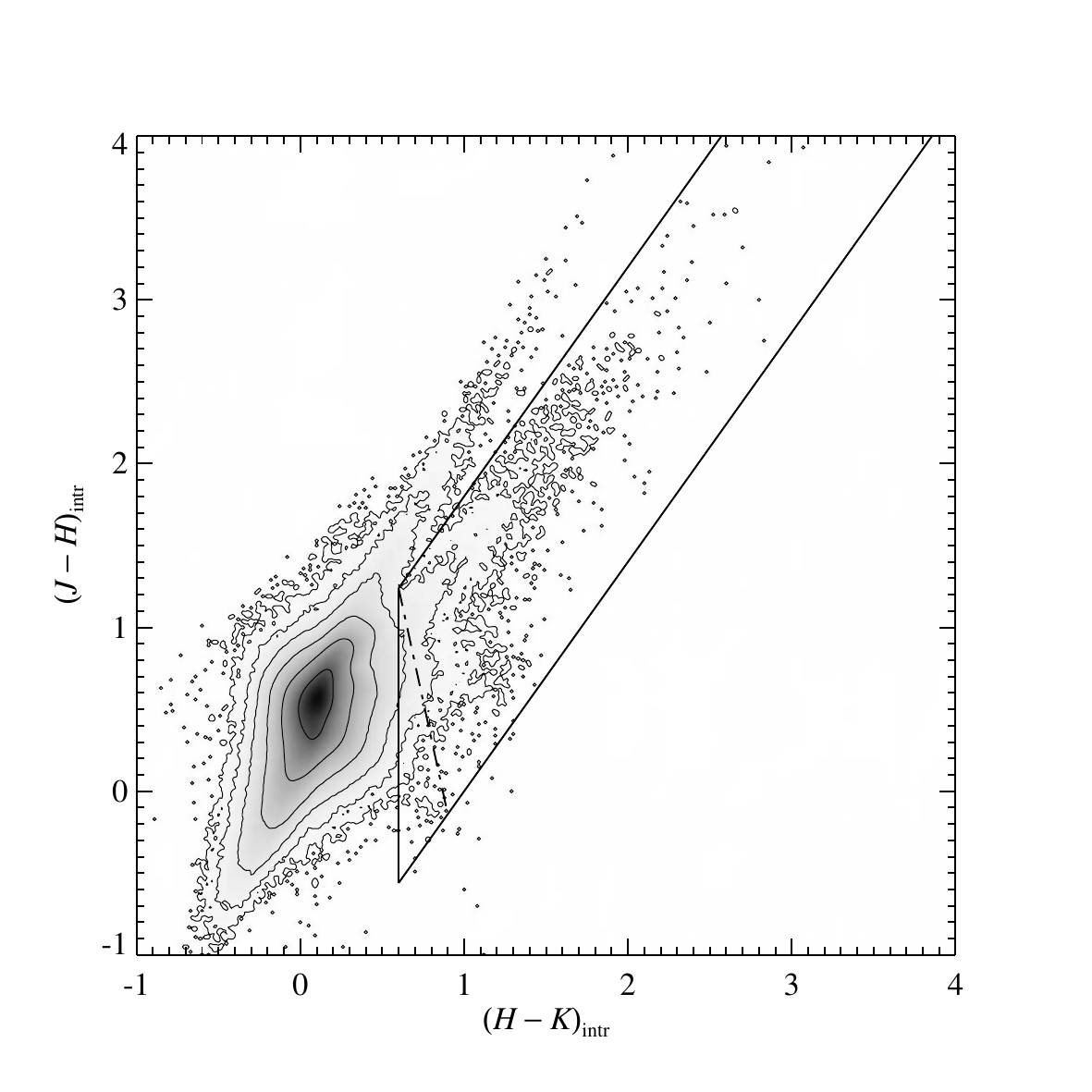}
  \caption{The extinction-corrected color-color diagram.  The stripes
    show both subsets $B_1$ and $B_2 \subset B_1$ [cf.\
    Eqs.~\eqref{eq:7} and \eqref{eq:8}].}
  \label{fig:4}
\end{figure}

\begin{figure}[!t]
  \centering
  \includegraphics[width=\hsize, bb=13 16 300 299]{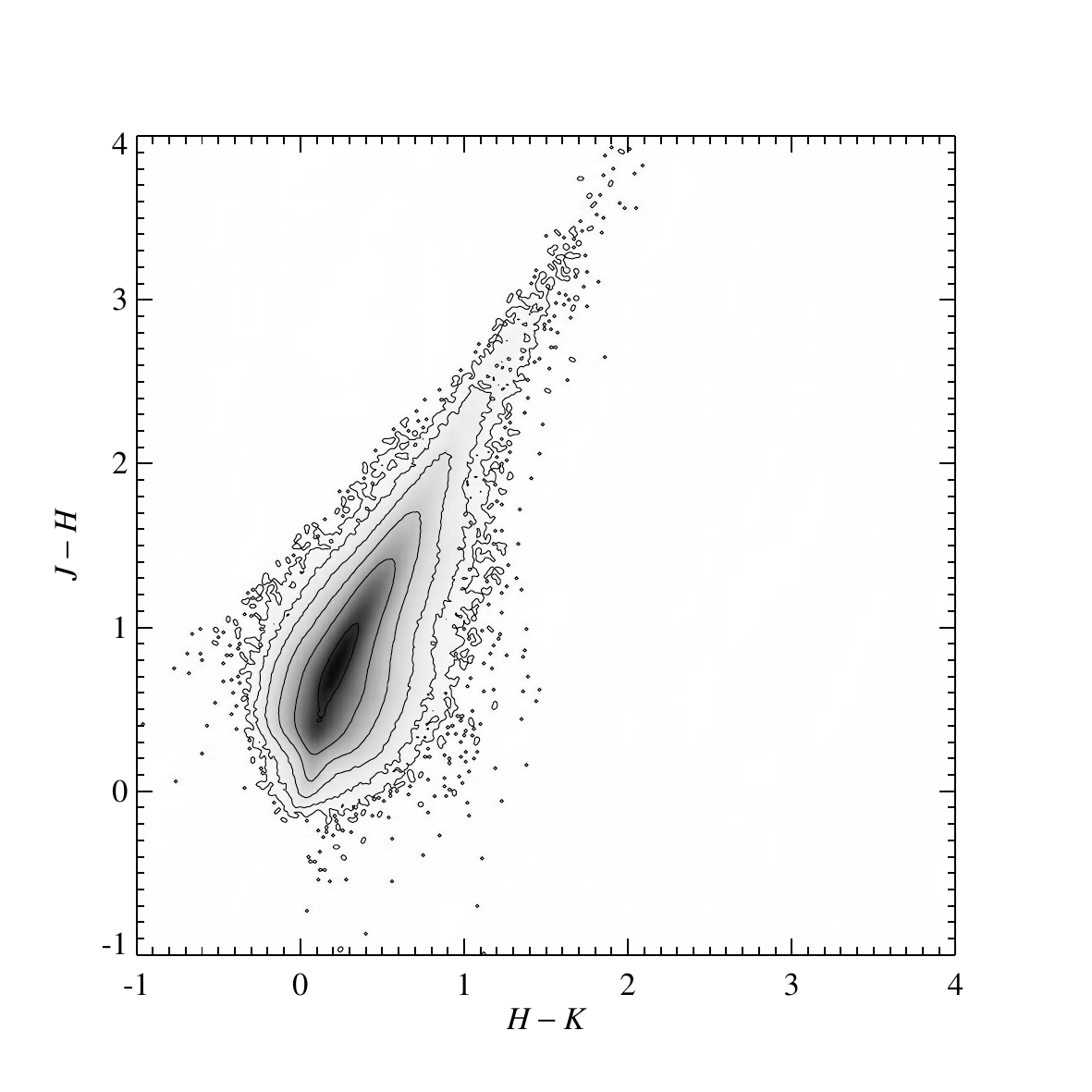}
  \caption{The color-color diagram for the selected stars in the
    field, after the removal of the set $B_2$ of Eq.~\eqref{eq:8}.  A
    comparison with Fig.~\ref{fig:1} shows that we were able to
    virtually remove all significant contamination from spurious
    reddening.}
  \label{fig:5}
\end{figure}

\begin{figure*}[!tbp]
  \begin{center}
    \includegraphics[width=\hsize]{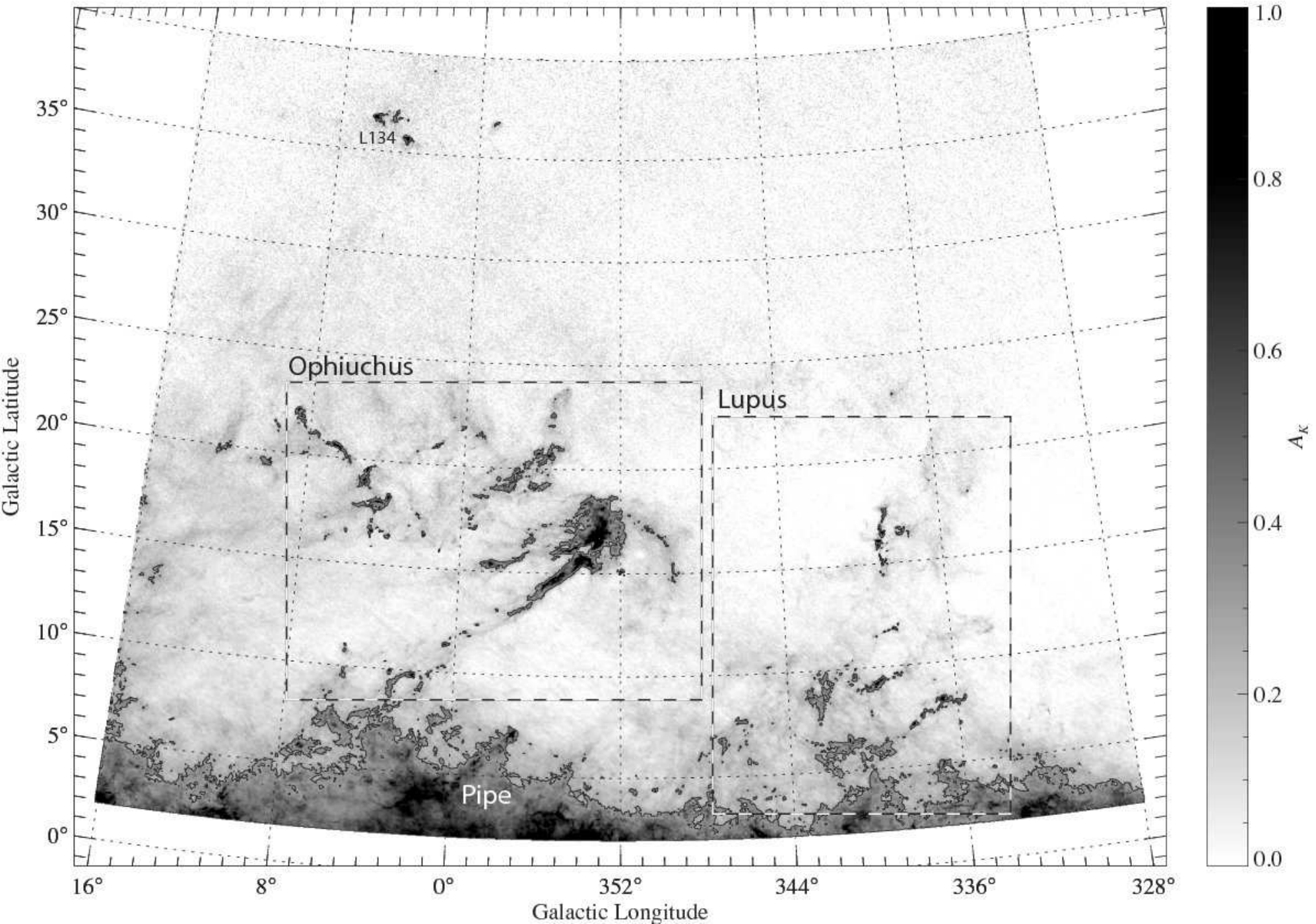}
    \caption{The \textsc{Nicer} extinction map of the Ophiuchus, Lupus,
      and Pipe complexes.  The resolution is $\mathrm{FWHM} = 3 \mbox{
        arcmin}$.  The various dashed boxes mark the regions shown in
      greater detail in Figs.~\ref{fig:9} and \ref{fig:10}.  The
      overplotted contour is at $A_K = 0.3 \mbox{ mag}$.  The Pipe
      nebula occupies the region around $(l, b) = (0^\circ, 5^\circ)$.}
    \label{fig:6}
  \end{center}
\end{figure*}

\begin{figure*}[!tbp]
  \begin{center}
    \includegraphics[width=\hsize]{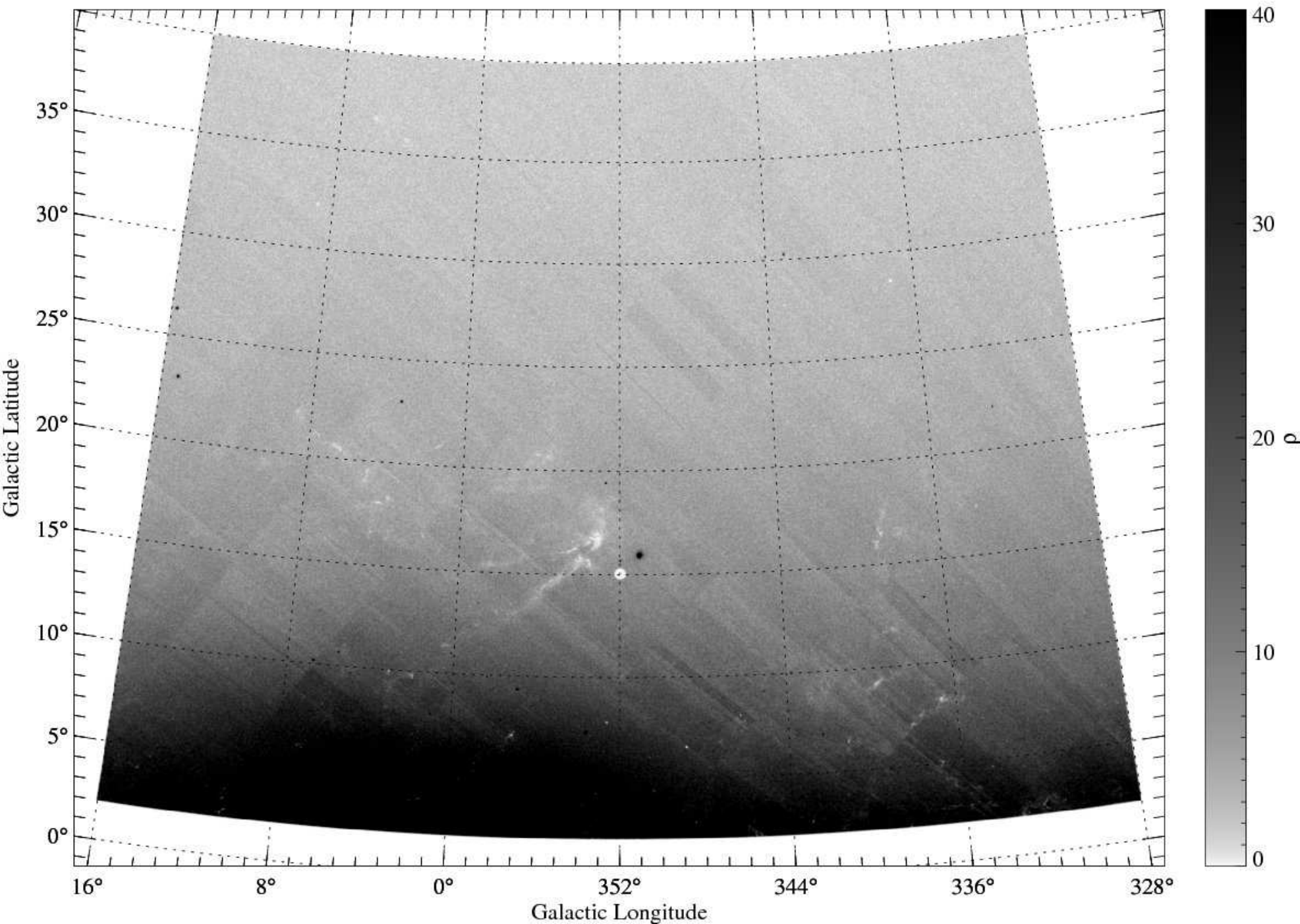}
    \caption{The star density for the extinction map of
      Fig.~\ref{fig:6}; note how evident is the galactic bulge here.
      The faint white structures correspond to the position of dense
      cores, where the star density decreases because of the cloud
      extinction.  The white ``hole'' at the center-bottom of the
      field is Antares.  The black spots are open and globular
      clusters.}
    \label{fig:7}
  \end{center}
\end{figure*}

\begin{figure*}[!tbp]
  \begin{center}
    \includegraphics[width=\hsize]{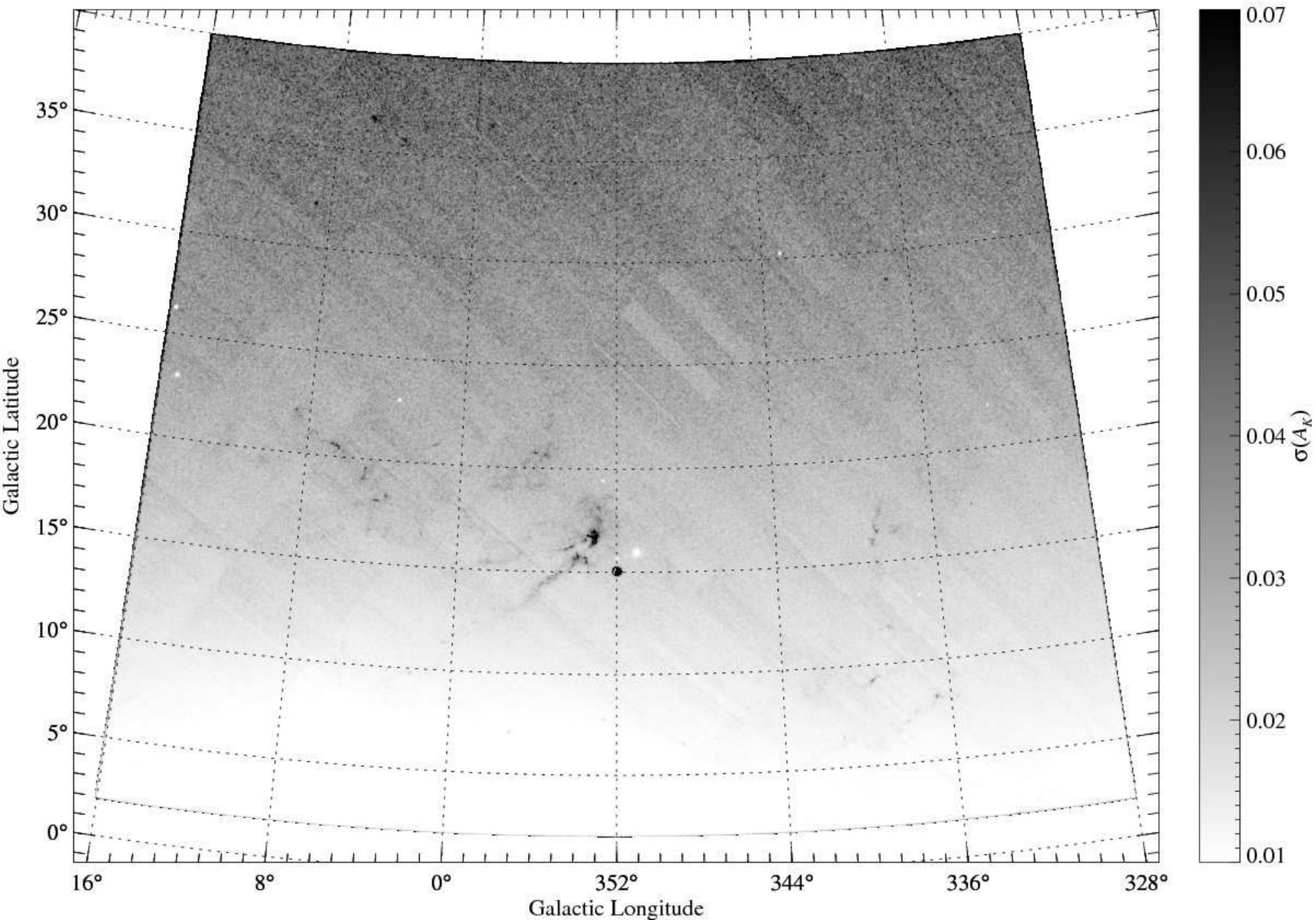}
    \caption{The statistical error $\sigma(A_K)$ on the
      measured column density of Fig.~\ref{fig:6}.  Note that
      correlation on the errors on a scale of $\mathrm{FWHM} = 3
      \mbox{ arcmin}$ is expected.  The increase of the error with the
      galactic latitude is evident.}
    \label{fig:8}
  \end{center}
\end{figure*}

\begin{figure*}[!tbp]
  \begin{center}
    \includegraphics[width=\hsize, bb=25 58 596 450]{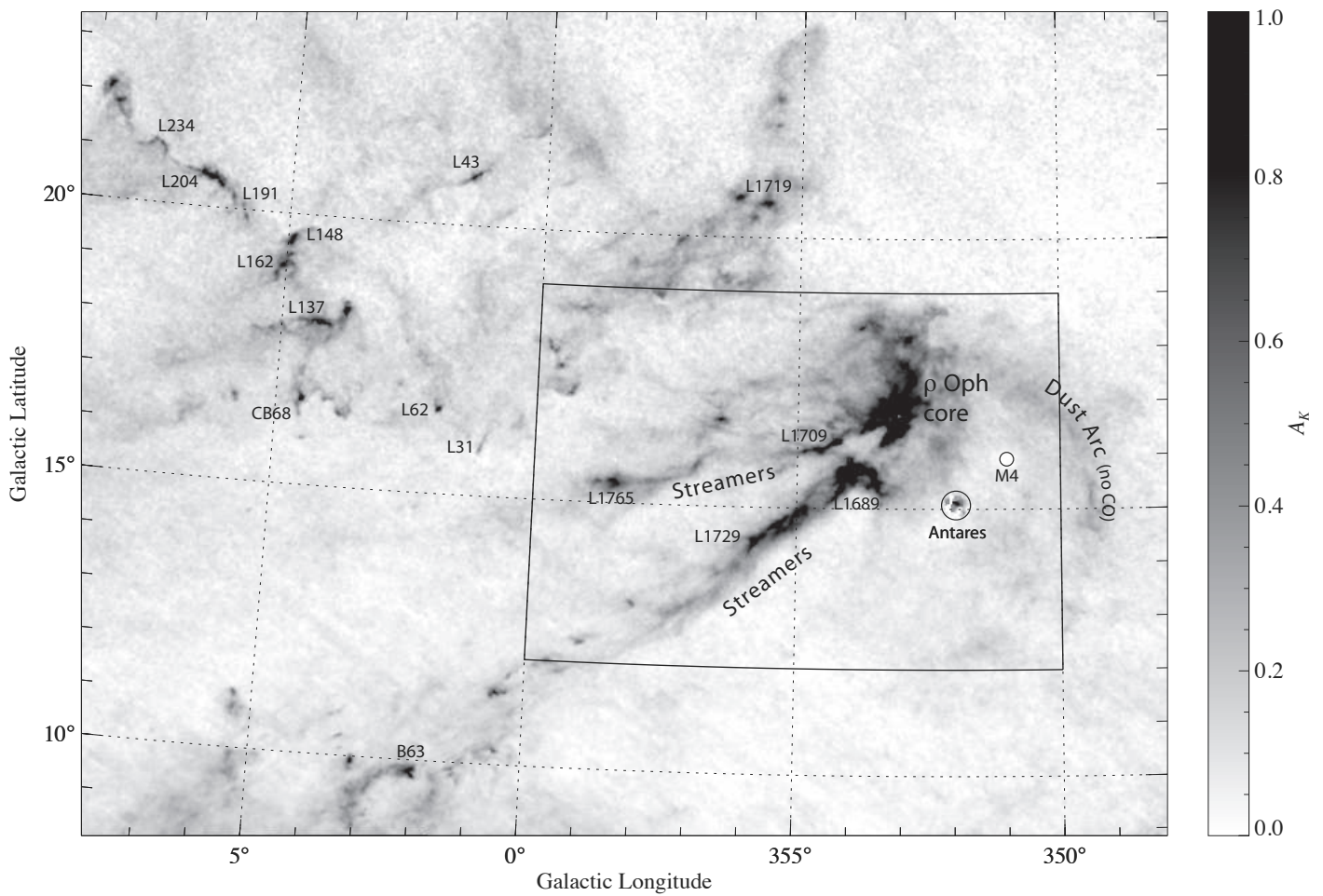}
    \caption{A zoom of Fig.~\ref{fig:6} showing the central region of
    the Ophiuchus cloud.  The box encloses the core of cloud, for which
    a few separate studies are presented in this paper.}
    \label{fig:9}
  \end{center}
\end{figure*}

\begin{figure*}[!tbp]
  \begin{center}
    \includegraphics[width=\hsize, bb=28 58 596 719]{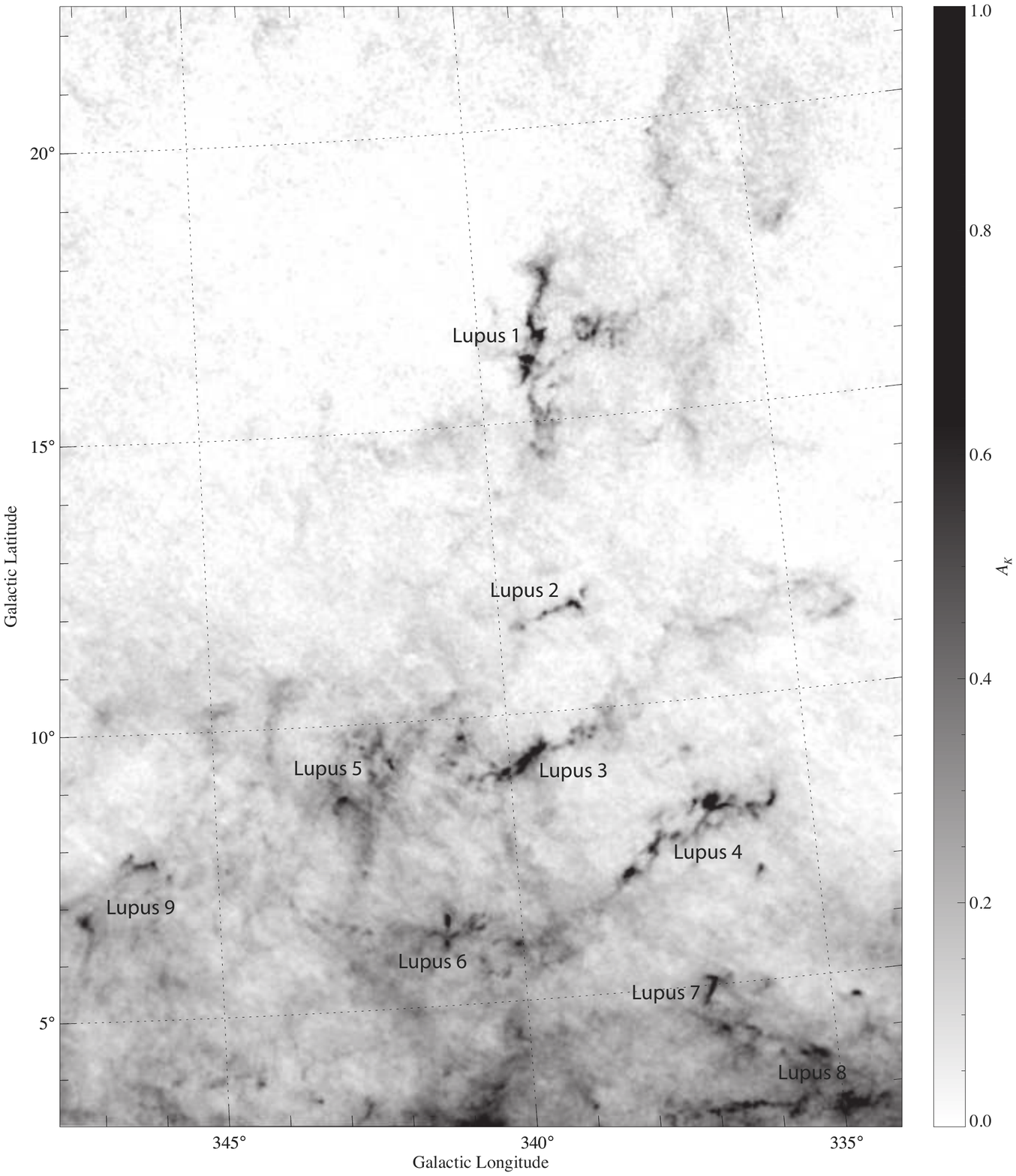}
    \caption{A zoom of Fig.~\ref{fig:6} showing the of Lupus cloud
      complex (in particular, all clouds from Lupus~1 to Lupus~9).}
    \label{fig:10}
  \end{center}
\end{figure*}

When selecting stars from the 2MASS point source catalog we discarded
possible spurious detections (e.g., objects likely to be associated to
solar system minor bodies, extended sources, or artifacts).  We then
generated a preliminary extinction map which, as described in Paper~I,
was mainly used as a first check of the parameters adopted, to select
a control region on the field, and to obtain the photometric
parameters to be used in the final map (see Fig.~\ref{fig:6}).  We
identified a large region that is apparently affected by only a
negligible extinction (see below), and used the colors of stars in
this control field as reference ones.

Using the information provided by the control field, we generated a
second map, which is thus ``calibrated'' (i.e., provides already, for
each position in our field, a reliable estimate of the column
density).  Note that the recent \citet{2005ApJ...619..931I} 2MASS
reddening law was used here.  We then considered the color-color
diagram for the stars in the catalog to check for possible signs of
anomalous star colors.  The result, presented in Fig.~\ref{fig:1}, shows a
bifurcation for $H - K > 1 \mbox{ mag}$.

As discussed in detail in Paper~II, the bifurcation is likely to be
created by Asymptotic Giant Branch (AGB) stars.  To further
investigate this point, we considered the two samples in the
color-color diagram defined as
\begin{align}
  \label{eq:2}
  A \equiv {} & \{ 1.4 (H - K) + 0.5 \mbox{ mag} < (J - H) \notag\\
  & \phantom{\bigl\{} \mbox{ and } H - K > 1 \mbox{ mag} \} \; , \\
  \label{eq:3}
  B \equiv {} & \{ 1.4 (H - K) + 0.5 \mbox{ mag} > (J - H) \notag\\
  & \phantom{\bigl\{} \mbox{ and } H - K > 1 \mbox{ mag} \} \; .
\end{align}
An analysis of the spatial distribution of these two samples
(Fig.~\ref{fig:2}) reveals that, as expected, sample $A$ is
associated with the densest regions of the molecular cloud, while
sample $B$ is distributed on the whole field with a strong
preference for low galactic latitude regions.

The nature of the two stellar populations in samples $A$ and $B$ is
further clarified by the histogram of their $K$ band magnitudes, shown
in Fig.~\ref{fig:3}.  As expected, sample $A$ shows a broad
distribution, which can be essentially described as a simple power-law
luminosity function up to $K \simeq 12 \mbox{ mag}$; note that the
completeness limit of our sample is significantly smaller than the
typical 2MASS completeness in the $K$ band ($14.3 \mbox{ mag}$ at
$99\%$ completeness) because of the stricter selection operated here
(small photometric errors in all bands) and because most sample $A$
stars come from low galactic latitude regions (where the increased
density of stars significantly reduces the completeness of the 2MASS).
In contrast to sample $A$ stars, sample $B$ stars show a well defined
distribution, with a pronounced and relatively narrow peak at $K
\simeq 7 \mbox{ mag}$.  This strongly suggests that we are looking at
a \textit{homogeneous\/} population of sources located at essentially
the same \textit{distance}.

The lack of correlation between the dust reddening and the stars of
sample $B$ can be also investigated by considering the
extinction-corrected color-color diagram shown in Fig.~\ref{fig:4}.
This plot was obtained by estimating, for each star, its ``intrinsic''
colors, i.e.\ the extinction corrected colors from the extinction at
the star's location as provided by the \textsc{Nicer} map.  In other
words, we computed
\begin{align}
  \label{eq:4}
  J_\mathrm{intr} & {} \equiv J - (A_J / A_K) \hat A_K \; , \\
  \label{eq:5}
  H_\mathrm{intr} & {} \equiv H - (A_H / A_K) \hat A_K \; , \\
  \label{eq:6}
  K_\mathrm{intr} & {} \equiv K - \hat A_K \; ,
\end{align}
where $\hat A_K$ is the \textsc{Nicer} estimated extinction in the
direction of the star from the angularly close objects.  By
comparing Fig.~\ref{fig:4} with Fig.~\ref{fig:1}, we see that the
upper branch, i.e. sample $A$, is strongly depressed, while the lower
branch (sample $B$) is largely unaffected.  Note that the residual
stars appearing in the upper branch are likely to be the effect of an
inaccurate extinction correction due to small-scale inhomogeneities
not captured by our analysis; similarly, the tail at negative colors
is due to ``over-corrected'' stars (for example foreground stars
observed in projection to a cloud).  Because of the much shrunken
distribution of the upper-branch, Fig.~\ref{fig:4} is particularly
useful to better identify stars belonging to the lower branch, and was
thus used to define two further subsets:
\begin{align}
  \label{eq:7}
  B_1 \equiv {} & \bigl\{ \bigl\lvert (J - H)_\mathrm{intr} - 1.4 (H -
  K)_\mathrm{intr} + 0.5 \bigr\rvert < 0.9 \notag\\ 
  & \phantom{\bigl\{} \mbox{ and } (J - H)_\mathrm{intr} > -4.6 (H -
  K)_\mathrm{intr} + 4 \} \; . \\
  \label{eq:8}
  B_2 \equiv {} & \bigl\{ \bigl\lvert (J - H)_\mathrm{intr} - 1.4 (H -
  K)_\mathrm{intr} + 0.5 \bigr\rvert < 0.9 \notag\\ 
  & \phantom{\bigl\{} \mbox{ and } (J - H)_\mathrm{intr} > -2.1 (H -
  K)_\mathrm{intr} + 2.5 \} \; .
\end{align} 
These two subsets, marked in Fig.~\ref{fig:4}, correspond to the areas in
the color-color plot where a contamination by sample $B$ stars is
possible ($B_1$), or highly likely ($B_2$).  We find approximately
$996\,000$ $B_1$ stars and $360\,000$ $B_2$ ones, corresponding to
$2.3\%$ and to $0.88\%$ respectively.

In summary, all the evidence found supports the identification of the
``lower branch'' with evolved intermediate mass stars (about 1 to 7
M$_\odot$), and likely with the Asymptotic Giant Branch (AGB) at about
the distance to the Galactic center (see also Paper~II).  Since the
``lower branch'' stars seem to be unrelated to the molecular cloud,
and since their colors would be interpreted by the \textsc{Nicer}
algorithm as a sign of extinction, in principle our results could be
biased toward a higher extinction, especially at low galactic latitude
regions.  In practice, we argue that the bias introduced by ``lower
branch'' stars is negligible given the low density of these stars.

Nevertheless, and in order to avoid any source of bias, although
small, we excluded from the 2MASS catalogs all stars located in the
$B_2$ color-space region, and performed the whole analysis described
in this paper using this reduced subset of stars.  We stress that if
we had performed a cut in the \textit{observed\/} colors, we would
have introduced a new bias in the deduced column density; instead, a
selection in the \textit{intrinsic\/} colors does not bias the final
results.  As an example, Fig.~\ref{fig:5} shows the color-color
diagram for the new set of stars: note that the ``lower branch''
disappears completely in this plot, a further confirmation that our
selection is effective in removing this population of stars.

We then run again the whole \textsc{Nicer} pipeline on the refined
catalog.  After (re)evaluating the statistical properties of stars in
the control field, we constructed the final map, shown in
Fig.~\ref{fig:6}.  We recall that in
\textsc{Nicer} the final map can be generated using different
smoothing techniques (see \citealp{2002A&A...395..733L} for a
discussion on the characteristics and merits of various
interpolators).  As pointed out in Paper~I, generally these
techniques produce comparable results, and thus we focused here to the
simple moving weight average:
\begin{equation}
  \label{eq:9}
  \hat A_K(\vec\theta) = \frac{\sum_{n=1}^N W^{(n)}(\vec\theta) \hat
    A^{(n)}_K }{\sum_{n=1}^N W^{(n)}} \; ,
\end{equation}
where $\hat A_K(\vec\theta)$ is the extinction at the angular position
$\vec\theta$, $\hat A^{(n)}_K$ is the extinction of the $n$-th star,
and $W^{(n)}(\vec\theta)$ is the weight for the $n$-th star for the
pixel at the location $\vec\theta$:
\begin{equation}
  \label{eq:10}
  W^{(n)}(\vec\theta) = \frac{W \bigl( \vec\theta - \vec\theta^{(n)}
    \bigr)}{\mathrm{Var}\bigl( \hat A^{(n)}_K \bigr)} \; .
\end{equation}
Hence, the weight for the $n$-th star is composed by two factors: (i)
$W \bigl( \vec\theta - \vec\theta^{(n)} \bigr)$, i.e.\ a function of
the angular distance between the star and the point $\vec\theta$ where
the extinction has to be interpolated, and (ii) $\mathrm{Var}\bigl(
\hat A^{(n)}_K \bigr)$, the inverse of the inferred variance on the
estimate of $A_K$ from the star.  The first factor, parametrized by
the weight function $W$, is here taken to be a Gaussian.

The map of Fig.~\ref{fig:6} was generated on a grid of approximately
$2\,000 \times 1\,600$ points, with scale $90 \mbox{ arcsec}$ per
pixel, and with Gaussian smoothing characterized by $\mbox{FWHM} = 3
\mbox{ arcmin}$.  Note that in the weighted average of
Eq.~\eqref{eq:9} we also introduced an iterative $\sigma$-clipping at
$3$-$\sigma$ error (see Paper~I).  The average, \textit{effective\/}
density of stars is $\sim 6.5$ stars per pixel, but as noted above
this value changes significantly on the field with the galactic
latitude (see Fig.~\ref{fig:7}); this density guarantees an average
($1$-$\sigma$) error on $A_K$ below $0.03$ magnitudes.  The largest
extinction was measured close to $\rho$ Ophiuchi, where $A_K \simeq
2.89$ magnitudes.  The expected error on the extinction $A_K$ is shown
in Fig.~\ref{fig:8}, and was evaluated from the relation (see Paper~II)
\begin{equation}
  \label{eq:11}
  \sigma^2_{\hat A_K}(\vec \theta) \equiv \frac{\sum_{n=1}^N \bigl[
    W^{(n)} (\vec\theta) \bigr]^2 \mathrm{Var} \bigl( \hat A_K^{(n)}
    \bigr)}{\Bigl[ \sum_{n=1}^N W^{(n)}(\vec\theta) \Bigr]^2} \; .
\end{equation}
As expected, we observe a significant gradient along the galactic
latitude.  Other variations in the expected errors can be associated
to bright stars (they produce the characteristic cross-shaped
patterns), to globular clusters (white dots in Fig.~\ref{fig:8}), and
to the cloud itself (dark areas).  Because of the relatively large
variations on the noise of the extinction map, clearly a detailed
analysis of Fig.~\ref{fig:6} should be carried out using in addition
the noise map of Fig.~\ref{fig:8}.  Figures~\ref{fig:9} and
\ref{fig:10} shows in greater detail the absorption maps we obtain for
the Ophiuchus and Lupus complexes, and allow us appreciate better the
details that we can obtain by applying the \textsc{Nicer} method to
the quality of 2MASS data.

\section{Statistical analysis}
\label{sec:statistical-analysis}

\subsection{Reddening law}
\label{sec:reddening-law}

\begin{figure}[!tbp]
  \begin{center}
    \includegraphics[width=\hsize, bb=16 7 331 240]{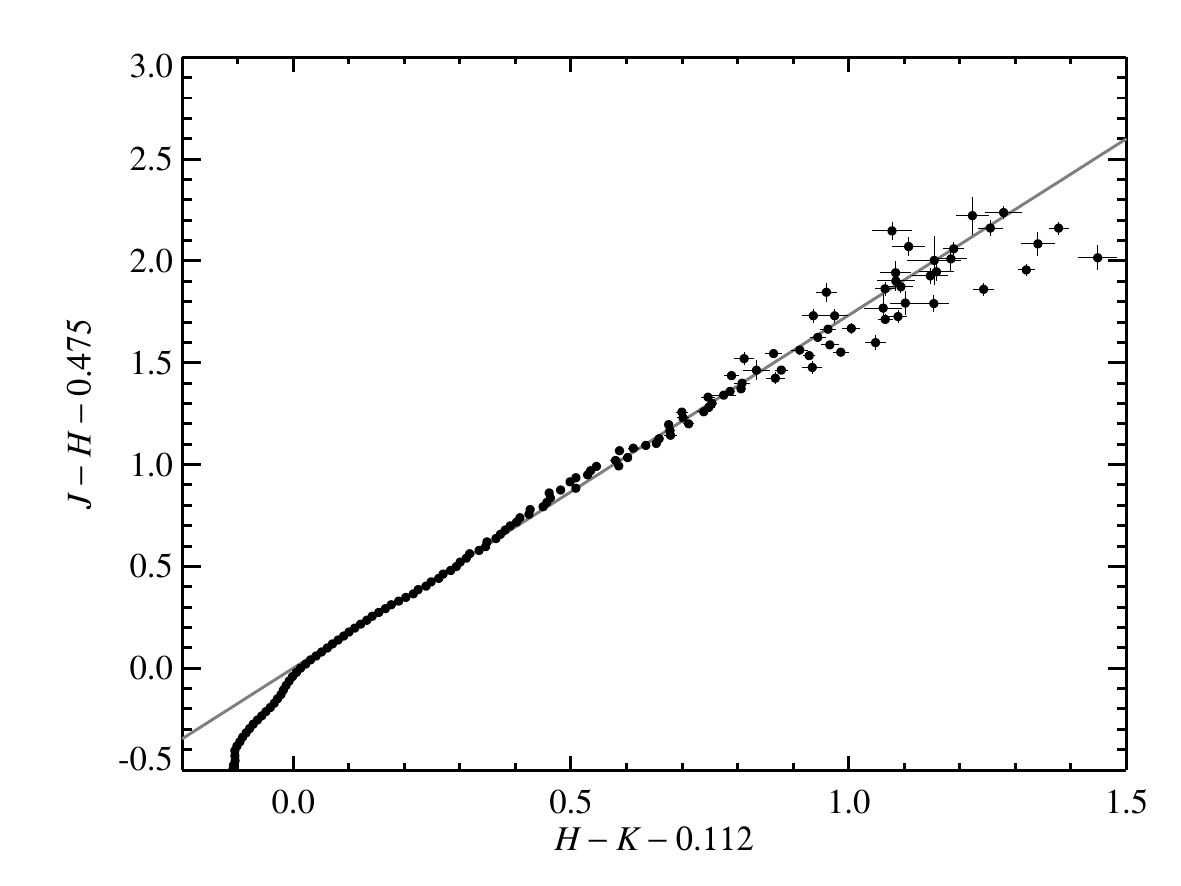}
    \caption{The reddening law as measured on the analyzed region.
      The plot shows the color excess on $J - K$ as a function of the
      color excess on $H - K$ (the constant $0.112$ and $0.475$
      represent, respectively, the average of $H - K$ and of $J - H$
      colors in magnitudes for the control field).  Error bars are
      uncertainties evaluated from the photometric errors of the 2MASS
      catalog.  The solid line shows the normal infrared reddening law
      \citep{2005ApJ...619..931I}.}
    \label{fig:13}
  \end{center}
\end{figure}

The large number of stars in the field allowed us to accurately check
the reddening law used throughout this paper.  To the purpose, we
partitioned all stars into different bins corresponding to the
individual original $\hat A_K^{(n)}$ measurements (we used a bin size of $0.02
\mbox{ mag}$).  Then, we evaluated the average NIR colors in each
group of stars in the same bin and the corresponding statistical
uncertainties (estimated from on the photometry errors of the 2MASS
catalog).  The results obtained are shown in Fig.~\ref{fig:13}
together with the normal infrared reddening law in the 2MASS
photometric system \citep{2005ApJ...619..931I}.  This plot shows that
there are no significant deviations from the normal reddening law over
the whole range of extinctions investigated here.

% \subsection{Photometric biases}
% \label{sec:photometric-biases}

% \begin{figure}[!tbp]
%   \begin{center}
%     \includegraphics[width=\hsize, bb=17 7 360 240]{fig14}
%     \caption{The average $\bigl\langle A_K^\mathrm{star} -
%       A_K^\mathrm{pixel} \bigr\rangle$ as a function of the star
%       absorption error (solid line, left scale).  Also shown here is
%       the histogram of the star absorption errors (right scale).}
%     \label{fig:xxx14}
%   \end{center}
% \end{figure}

% Figure~\ref{fig:xxx14} shows the average $\bigl\langle A_K^\mathrm{star}
% - A_K^\mathrm{field} \bigr\rangle$ (i.e.\ the difference between the
% extinction estimate for a single star and the average estimate in the
% corresponding part of the sky, deduced from the map of
% Fig.~\ref{fig:6}) as a function of the star absorption error.  As
% shown by this plot, stars with poor photometry, and thus with a large
% estimated error $\mathrm{Err}\bigl( \hat A_K^{(n)} \bigr)$, provide
% column density estimates brighter than the ones obtained from better
% measured stars.  Although the interpretation of this result is not
% straightforward, we take it as an indication that the colors of stars
% with large photometric errors could be biased toward the red.  Hence,
% for robustness, all stars with $\mathrm{Err}\bigl( \hat A_K^{(n)}
% \bigr) > 0.18 \mbox{ mag}$ have been rejected in our analysis.  We
% also note that this bias might affect the determined reddening (cf.\
% Sect.~\ref{sec:reddening-law}) and could be responsible for the
% apparent change of slope at large column densities visible in
% Fig.~\ref{fig:13}.

\subsection{Foreground star contamination}
\label{sec:foregr-star-cont}

\begin{figure}[!tbp]
  \begin{center}
    \includegraphics[width=\hsize]{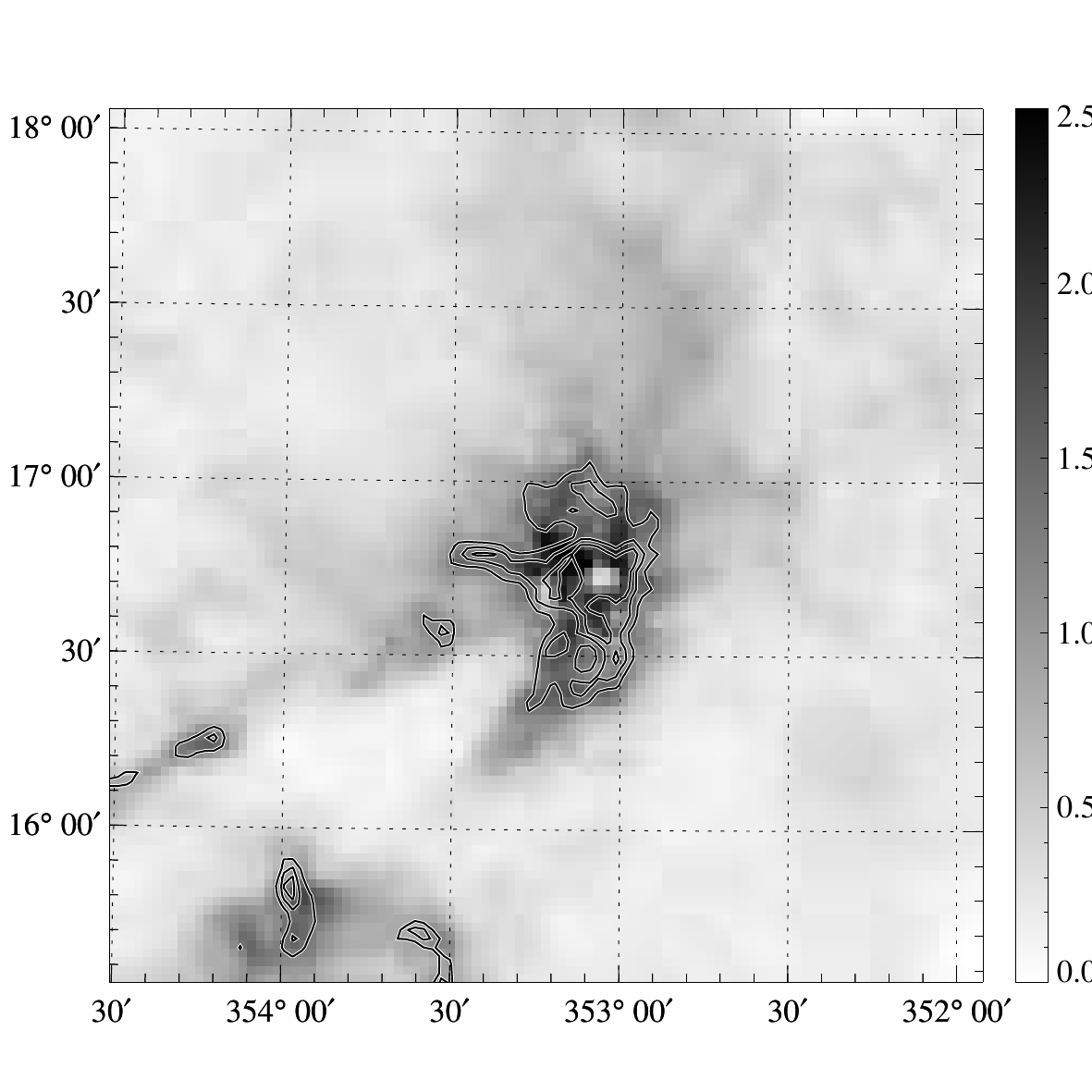}
    \caption{A zoom of the extinction map near $\rho$ Ophiuchi, with
      overprinted the star density contours.  Note the white ``hole''
      close to the center of this core, which is due to the combined
      effect of very high extinction values and the presence of
      embedded stars in this star forming core.}
    \label{fig:14}
  \end{center}
\end{figure}

If a fraction $f$ of observed stars is foreground to a dark cloud, the
measured column density is underestimated by a factor $(1 - f)$.  This
effect is normally negligible on the outskirts of nearby molecular
clouds, where the fraction $f$ of foreground stars is typically of the
order $\sim 1\%$, but unfortunately can have significant effects on
very dense regions, where because of a selection effect $f$ increases
significantly.  In addition, many dense cores host young stellar
objects: these stars, if moderately embedded, show only a fraction of
the true, total column density of the cloud.  As a result, the
extinction in the direction of dense regions can be severely
underestimated.  For example, the core of $\rho$ Ophiuchi shows an
apparent ``hole'' in absorption (see Fig.~\ref{fig:14}), but a
comparison with the local density of stars shows that the hole is the
result of young stars moderately embedded on this active star forming
region (this core host $\sim 200$ young stellar objects) and of
foreground star contamination (see also \cite{1997ApJS..112..109B} for
a discussion on the effects of foreground stars in the $\rho$ Ophiuchi
core).

In order to evaluate quantitatively the fraction $f$ of foreground
stars, we selected high-extinction regions characterized by $A_K > 1
\mbox{ mag}$.  We then checked all stars in these regions that show
``anomalous'' extinction, i.e.\ stars whose column densities differ by
more than 3-$\sigma$ with respect to the field.  A total of $1\,268$
stars met this criteria, but only $784$ of them show measured column
densities compatible with no extinction.
%@@@ CHECK THIS!
Hence, since the total area of regions with $A_K > 1 \mbox{ mag}$ is
about $1.4 \mbox{ deg}^2$, we estimate that on average on the field
only a fraction $\sim 2 \%$ of stars is foreground.  As a result, we
can safely ignore the effect of foreground stars except on the higher
extinction regions which, however, are not the focus of this paper.
%@@@ DISCUSS MAX A_K

\subsection{Column density probability distribution}
\label{sec:column-dens-prob}

\begin{figure}[!tbp]
  \begin{center}
    \includegraphics[width=\hsize, bb=22 7 331 183]{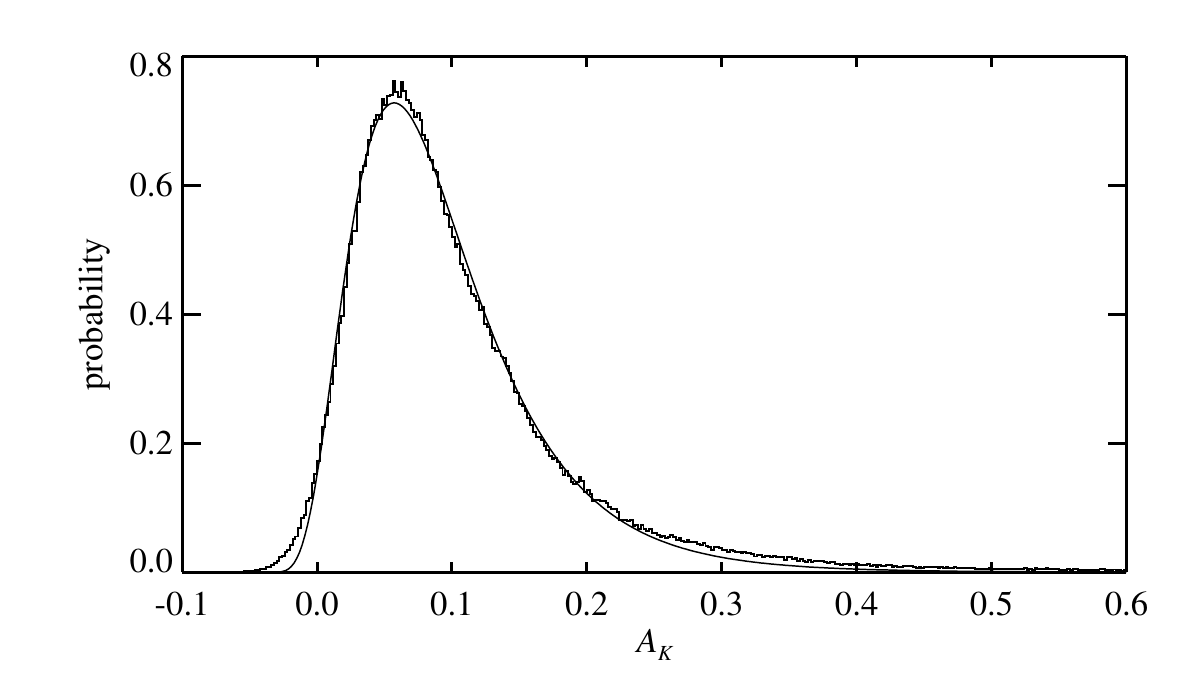}
    \caption{The probability distribution of star pixel extinctions
      for the Ophiuchus map shown in Fig.~\ref{fig:9}; the gray, smooth
      curve represents the best-fit with a log-normal distribution.}
    \label{fig:11}
  \end{center}
\end{figure}

\begin{figure}[!tbp]
  \begin{center}
    \includegraphics[width=\hsize, bb=22 7 331 183]{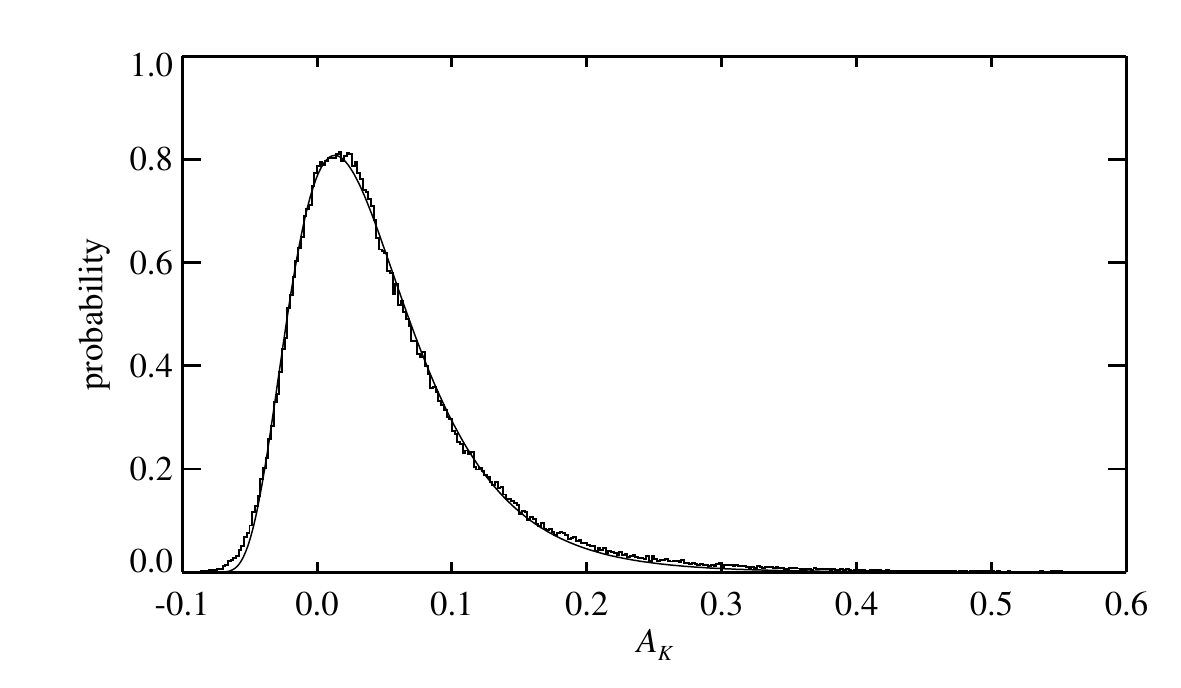}
    \caption{The probability distribution of star pixel extinctions
      for the subset $b > 6.5^\circ$) of the Lupus region of
      Fig.~\ref{fig:10}; the gray, smooth curve represents the
      best-fit with a log-normal distribution.}
    \label{fig:12}
  \end{center}
\end{figure}

\begin{table}[b!]
  \centering
  \begin{tabular}{lcccc}
    & Center $A_0$ & Scale $A_1$ & Dispersion $\sigma$ & Peak $a$ \\
    \hline
    Ophiuchus               & $-0.081$ & $0.151$ & $0.681$ & $946$ \\
    Lupus ($b > 6.5^\circ$) & $-0.135$ & $0.153$ & $1.506$ & $651$ 
  \end{tabular}
  \caption{The best-fit parameters of the four Gaussian functions used
    to fit the column density probability distribution shown in
    Figs.~\ref{fig:11} and \ref{fig:12} (see Eq.~\eqref{eq:12} for the
    meaning of the various quantities).}
  \label{tab:1}
\end{table}

Theoretical studies \citep[e.g.][]{1994ApJ...423..681V,
  1997MNRAS.288..145P, 1998PhRvE..58.4501P, 1998ApJ...504..835S} have
shown that the probability distribution for the \textit{volume
  density\/} in molecular clouds is log-normal for
\textit{isothermal\/} flows (i.e., when the polytropic index $\gamma
\equiv \diff \ln P / \diff \ln \rho = 1$), while it tends to develop a
power-law tail at high (respectively, low) densities for $\gamma < 1$
($\gamma > 1$).  In reality, observations can only probe the
probability distribution for the \textit{column density}, i.e.\ the
volume density integrated along the line of sight.  As discussed by
\citet{2001ApJ...557..727V}, this quantity can behave differently
depending on the ratio $\eta$ of the column density to a suitable
defined ``decorrelation length.''  In particular, if $\eta$ is large,
so that there are many independent ``events'' along the line of sight,
the probability distribution for the column density is expected to be
normal, a result in agreement with the central limit theorem
\citep[see, e.g.,][]{Eadie}; if, instead, $\eta$ is small, the
probability distribution for the column density is expected to be
similar to the one for the volume density, i.e.\ log-normal in the
isothermal case.  In practice, numerical simulations show that the
convergence to the normal distribution is extremely slow and that
transition between the two cases, in typical cases, is for $\eta
\simeq 10^4$ \citep{2001ApJ...557..727V}.

Figures~\ref{fig:11} and \ref{fig:12} report the probability
distributions of column densities observed in the Ophiuchus and Lupus
fields, i.e.\ the relative probability of column density measurements
for each pixel of Figs.~\ref{fig:9} and \ref{fig:10}.  We tried to fit
the column density histograms with log-normal distributions of the
form
\begin{equation}
  \label{eq:12}
  h(A_K) = \frac{a}{A_K - A_0} 
  \exp\left[- \frac{\bigl(\ln (A_K - A_0) - \ln A_1 \bigr)^2}%
    {2 (\ln \sigma)^2} \right] \; .
\end{equation}
For the Ophiuchus complex we obtained a satisfactory fit on the whole
region; for Lupus, the fit is extremely good, especially when
excluding the low-galactic latitude regions ($b < 6.5^\circ$), which
are likely to be contaminated by different cloud complexes.  Note, by
comparison, that as described in Paper~I, for the Pipe nebula a good
fit to the column density distribution requires four normal
distribution.  This result might indicate that the Pipe nebula is the
result of the superposition of different components, each of which is
likely to be extended on the line of sight (so that the central limit
theorem can be applied to the total measured column density).

\subsection{Small-scale inhomogeneities}
\label{sec:small-scale-inhom}

It has been long recognized \citep{1994ApJ...429..694L} that the local
dispersion of extinction measurements increases with the column
density: in other words, for a fixed (small) patch of the sky, the scatter
of the individual star estimates of $A_K$ increases as the average of
$A_K$ increases.  The scatter in the $A_K$ estimates is typically
associated with the intrinsic scatter in the NIR star colors and to
the effect of small-scale inhomogeneities in the cloud projected
density.  These inhomogeneities are important for a number of factors:
\begin{itemize}
\item In the simplest interpretation, they indicates differential
  extinction or strong gradients in the column density (with typical
  scales smaller than the resolution of the extinction map).
\item If present at different scales, they can be interpreted as the
  effects of turbulent motions \cite[see, e.g.][]{1994ApJ...429..645M,
    1997ApJ...474..730P}.
\item In presence of significant inhomogeneities at scales smaller
  than the resolution of the extinction map, \textsc{Nicer} (as well
  as other color-excess methods) is expected to be biased toward low
  extinction.  This happens because the background stars will no
  longer be randomly distributed in the patch of the sky used to
  estimate the local extinction value, but will be preferentially
  detected in low-extinction regions (see
  \citealp{2005A&A...438..169L} for a more detailed discussion on this
  point).
\end{itemize}

In order to better quantify the effect of inhomogeneities on small
scales, consider the quantity (cf.\ Paper~II)
\begin{equation}
  \label{eq:13}
  \hat\sigma^2_{\hat A_K}(\vec\theta) = \frac{\sum_{n=1}^N W^{(n)} 
    \bigl[ \hat A_K^{(n)} - \hat A_K(\vec\theta)
    \bigr]^2}{\sum_{n=1}^N W^{(n)}} \; .
\end{equation}
Note that $\hat\sigma^2_{\hat A_K}$ is defined in a different way with
respect to $\sigma^2_{\hat A_K}$ of Eq.~\eqref{eq:11}.  Let us now fix
a given direction in the sky $\vec\theta$, and let us consider the
process of measuring the column density $\hat A_K(\vec\theta)$ there.
This quantity is evaluated using Eq.~\eqref{eq:9}, i.e.\ $\hat
A_K(\vec\theta)$ is a weighted mean of the estimated column densities
$\bigl\{ \hat A_K^{(n)} \bigr\}$ of the stars observed close to the
direction $\vec\theta$.  The estimated star column density for the
$n$-th star, in turn, can be written as
\begin{equation}
  \label{eq:14}
  \hat A^{(n)}_K = \tilde A_K + \Delta^{(n)} + \epsilon^{(n)} \; .
\end{equation}
In this equation we split the three contributions to the measured
column density: $\tilde A_K$, the ``average'' extinction in the patch
of the sky considered (see below); $\Delta^{(n)} \equiv A_K\bigl(
\theta^{(n)} \bigr) - \tilde A_K$, the local difference from the
average extinction, which includes both random inhomogeneities (e.g.,
due to turbulence) and unresolved structures (e.g., due to steep
gradients in the extinction); and $\epsilon^{(n)}$, the photometric
error on the measured extinction of the $n$-th star.

Because of the presence of the photometric error $\epsilon^{(n)}$ in
Eq.~\eqref{eq:14}, we cannot use directly the estimator
$\hat\sigma^2_{\hat A_K}$ of Eq.~\eqref{eq:13} as a measure of
small-scale inhomogeneities, and a more detailed analysis is needed.
In the following we will consider \textit{ensemble averages\/} with
respect to the photometric errors $\epsilon^{(n)}$; in other words, we
will evaluate the mean values and variances of some relevant
quantities by taking $\epsilon^{(n)}$ as independent random variables
with the properties
\begin{align}
  \label{eq:15}
  \bigl\langle \epsilon^{(n)} \bigr\rangle = {} & 0 \; , &
  \bigl\langle (\epsilon^{(n)} )^2 \bigr\rangle = {} &
  \mathrm{Var}\bigl( \hat A_K^{(n)} \bigr) \equiv V^{(n)} \; , &
\end{align}
where the last equality is a mere definition.  In the decomposition of
Eq.~\eqref{eq:14} we can freely choose the value of the ``average''
extinction $\tilde A_K$ (a change of $\tilde A_K$ can be
counterbalanced by a change of all $\Delta^{(n)}$); we now fix
this quantity to
\begin{equation}
  \label{eq:16}
  \tilde A_k \equiv \frac{\sum_{n=1}^N W^{(n)}
    A_K\bigl(\vec\theta^{(n)}\bigr)}{\sum_{n=1}^N W^{(n)}} \; .
\end{equation}
This is a convenient and natural choice since in this case we have
\begin{equation}
  \label{eq:17}
  \frac{\sum_{n=1}^N W^{(n)} \Delta^{(n)}}{\sum_{n=1}^N W^{(n)}} = 0
  \; ,
\end{equation}
and $\bigl\langle \hat A_K(\vec\theta) \bigr\rangle = \tilde A_K$.
Note also that in this case we have
\begin{equation}
  \label{eq:18}
  \hat A_K(\vec\theta) - \tilde A_K = \frac{\sum_{n=1}^N W^{(n)}
    \epsilon^{(n)}}{\sum_{n=1}^N W^{(n)}} \equiv \epsilon(\vec\theta)
  \; .
\end{equation}
The quantity $\epsilon(\vec\theta)$ can be interpreted as the
``error'' on the extinction map:
\begin{align}
  \label{eq:19}
  \bigl\langle \epsilon(\vec\theta) \bigr\rangle = {} & 0 \; , \\
  \label{eq:20}
  \bigl\langle \bigl[ \epsilon(\vec\theta) \bigr]^2 \bigr\rangle = {} &
  \frac{\sum_{n=1}^N \bigl( W^{(n)} \bigr)^2 V^{(n)} }{\Bigl(
    \sum_{n=1}^N W^{(n)} \Bigr)^2} \equiv \sigma^2_{\hat A_k}(\vec\theta) \; ,
\end{align}
where in the last step we used the definition \eqref{eq:11}.

We can now finally consider the ensemble average of the quantity of
Eq.~\eqref{eq:13}:
\begin{equation}
  \label{eq:21}
  \bigl\langle \hat\sigma^2_{\hat A_K} \bigr\rangle  =
  \frac{\sum_{n=1}^N W^{(n)} \Bigl\langle \bigl[ \Delta^{(n)} +
    \epsilon^{(n)} - \epsilon(\vec\theta) \bigr]^2 
    \Bigr\rangle}{\sum_{n=1}^N W^{(n)}} \; . 
\end{equation}
Using the relation
\begin{equation}
  \label{eq:22}
  \bigl\langle \epsilon^{(n)} \epsilon(\vec\theta) \bigr\rangle = 
  \left\langle \epsilon^{(n)} \frac{\sum_m W^{(m)}
      \epsilon^{(m)}}{\sum_m W^{(m)}} \right\rangle = \frac{W^{(n)}
    V^{(n)}}{\sum_m W^{(m)}} \; ,
\end{equation}
we can expand Eq.~\eqref{eq:21} into
\begin{align}
  \label{eq:23}
  \bigl\langle \hat\sigma^2_{\hat A_K} \bigr\rangle = {} &
  \frac{\sum_n W^{(n)} \bigl( \Delta^{(n)} \bigr)^2}{\sum_n W^{(n)}} +
  \frac{\sum_n W^{(n)} V^{(n)}}{\sum_n W^{(n)}} + 
  \sigma^2_{\hat A_K}(\vec\theta) \notag\\
  & {} - 2 \frac{\sum_n \bigl( W^{(n)} \bigr)^2 V^{(n)}}{\Bigl( \sum_n
    W^{(n)} \Bigr)^2} \notag\\
  {} = {} &
  \frac{\sum_n W^{(n)} \bigl( \Delta^{(n)} \bigr)^2}{\sum_n W^{(n)}} +
  \frac{\sum_n W^{(n)} V^{(n)}}{\sum_n W^{(n)}} -
  \sigma^2_{\hat A_K}(\vec\theta) \; .
\end{align}
The first term in the r.h.s.\ of this equation represents a measure of
small scale inhomogeneities.  Since all other quantities appearing in
Eq.~\eqref{eq:23} can be evaluated from the data, it makes sense to
define the map
\begin{align}
  \label{eq:24}
  \Delta^2(\vec\theta) & {} \equiv \hat\sigma^2_{\hat A_K}(\vec\theta) +
  \sigma^2_{\hat A_K}(\vec\theta) - \frac{\sum_n W^{(n)}
    V^{(n)}}{\sum_n W^{(n)}} \notag\\ 
  & {} = \frac{\sum_n W^{(n)} \bigl( \Delta^{(n)} \bigr)^2}{\sum_n
    W^{(n)}} \; ,
\end{align}
and to interpret it as a ``variance'' of small scale inhomogeneities.
Note finally that if the weights $\bigl\{ W^{(n)} \bigr\}$ are chosen
according to Eq.~\eqref{eq:10}, then in the numerator of the last term
of this equation we can use $W^{(n)} V^{(n)} = W\bigl( \vec\theta -
\vec\theta^{(n)} \bigr)$.

\begin{figure}[!tbp]
  \begin{center}
    \includegraphics[width=\hsize]{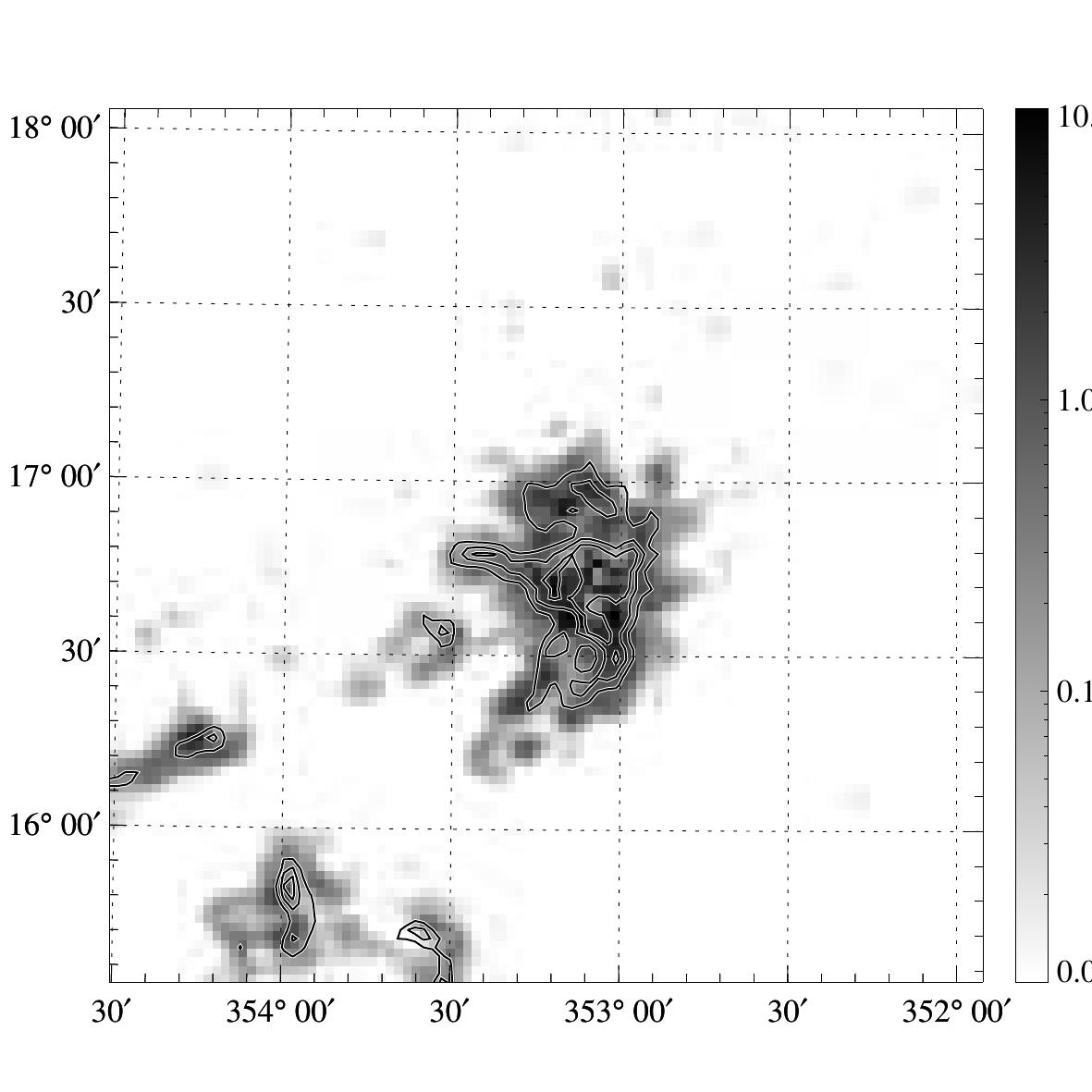}
    \caption{The $\Delta^2$ map of Eq.~\eqref{eq:24} on the same
      region around $\rho$ Ophiuchi shown in Fig.~\ref{fig:14}, with
      overplotted star density contours.  Note the significant
      increase observed in $\Delta^2$ close to the central parts of
      this cloud.}
    \label{fig:15}
  \end{center}
\end{figure}

\begin{figure}[!tbp]
  \begin{center}
    \includegraphics[width=\hsize]{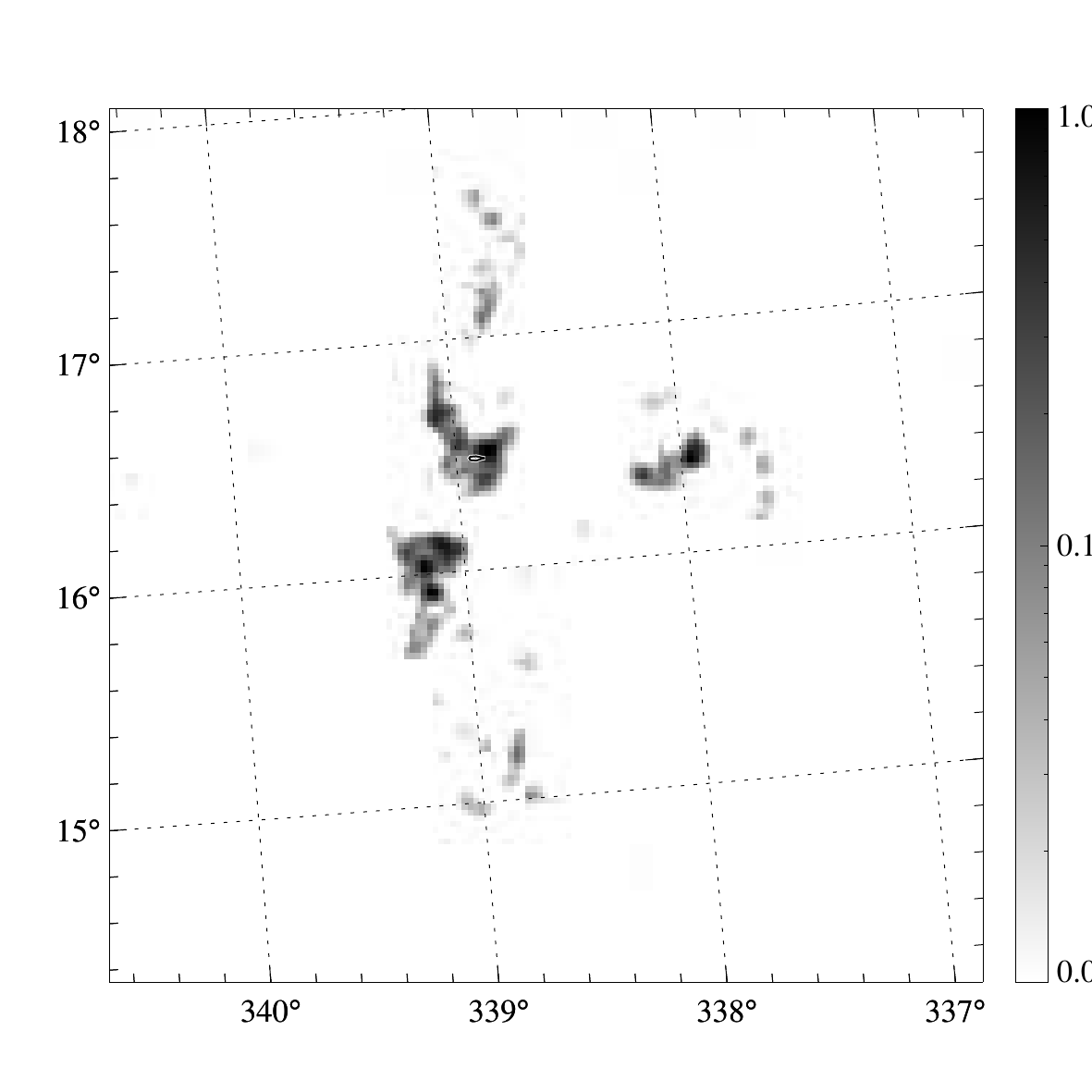}
    \caption{The $\Delta^2$ map of Eq.~\eqref{eq:24} on a region close
      to Lupus~1, in logarithmic scale, with overplotted start density
      contours (using the same levels as Fig.~\ref{fig:15}).}
    \label{fig:16}
  \end{center}
\end{figure}

\begin{figure}[!tbp]
  \begin{center}
    \includegraphics[bb=16 7 325 212, width=\hsize]{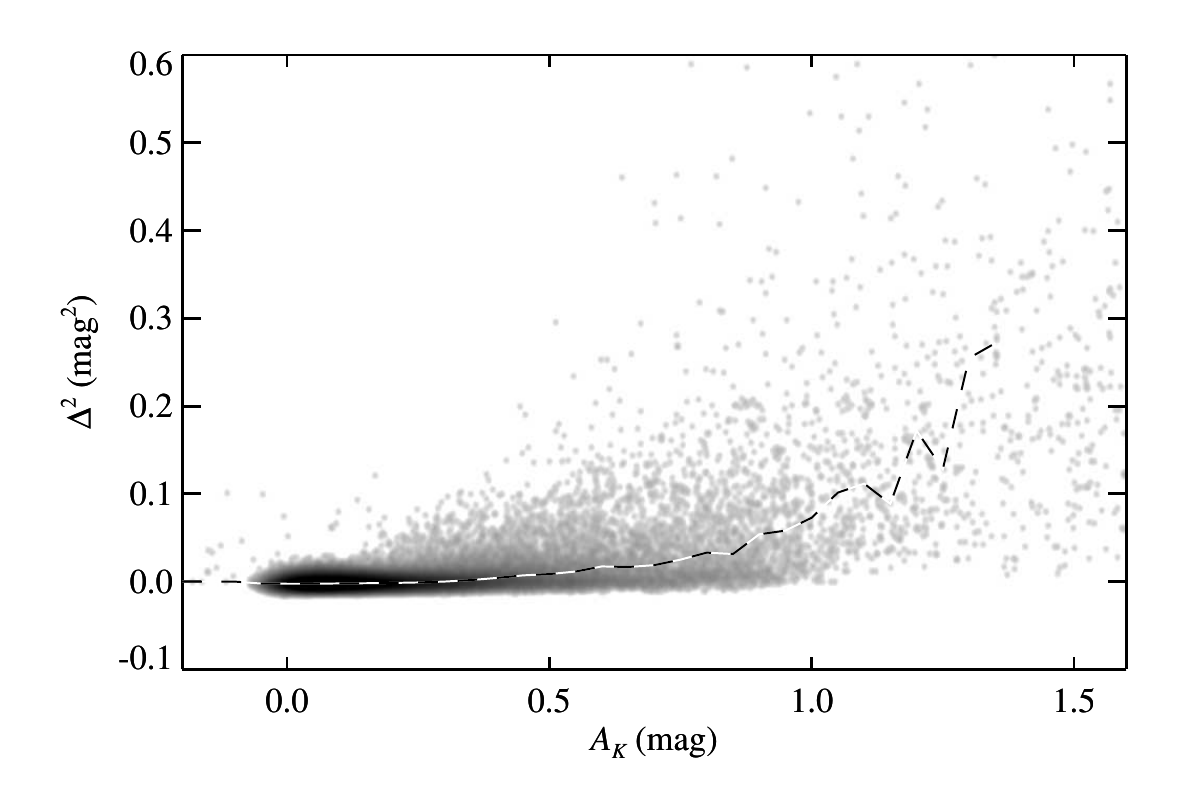}
    \caption{The distribution of the $\Delta^2$ map as a function of
      the local extinction $A_K$ for the Ophiuchus, in logarithmic
      grey scale.  The dashed line shows the average values of
      $\Delta^2$ in bins of $0.05 \mbox{ mag}$ in $A_K$.  Note the
      rapid increase of $\Delta^2$ for $A_K > 0.8 \mbox{ mag}$.  As a
      comparison, the average variance $\mathrm{Var}\bigl( \hat
      A^{(n)}_K \bigr)$ on the estimate of $\hat A_K$ from a single
      star is approximately $0.033 \mbox{ mag}^2$.}
    \label{fig:17}
  \end{center}
\end{figure}

\begin{figure}[!tbp]
  \begin{center}
    \includegraphics[bb=20 7 331 212, width=\hsize]{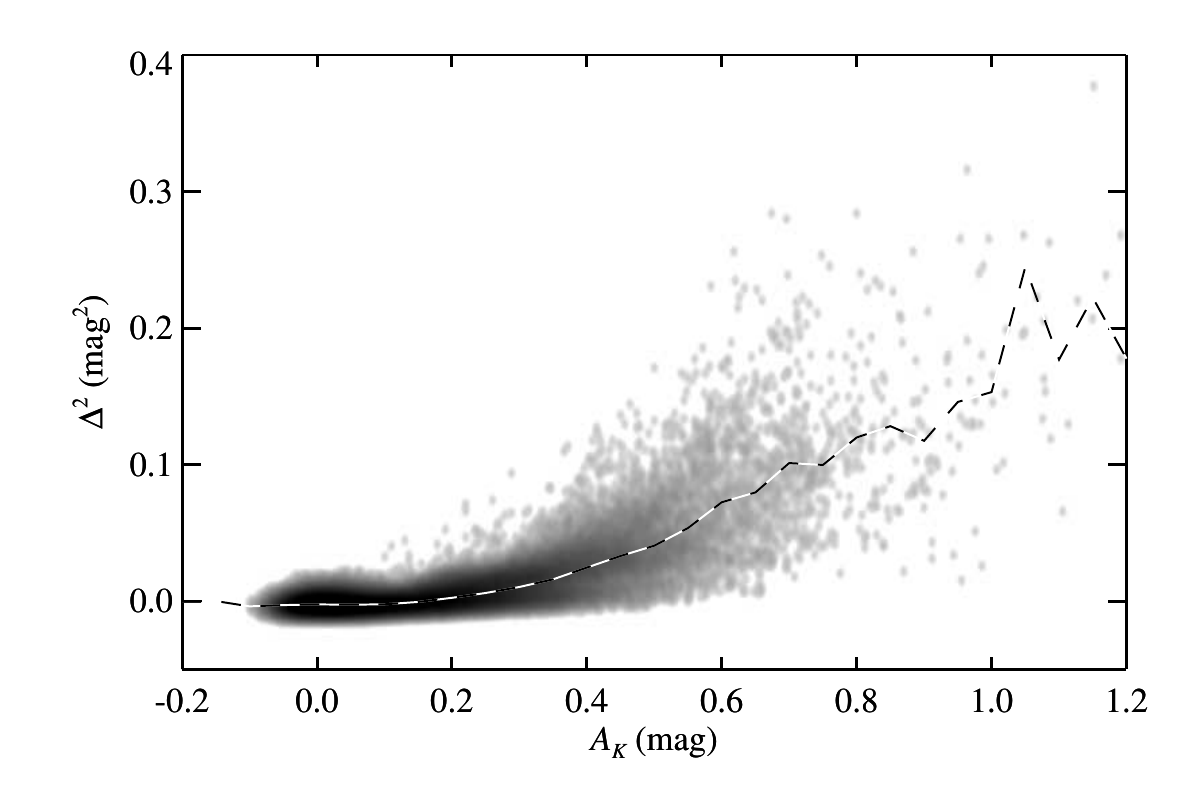}
    \caption{Same as Fig.~\ref{fig:17} for Lupus.  Note that the
      effect of substructures is evident at significantly lower
      column densities.}
    \label{fig:18}
  \end{center}
\end{figure}

We evaluated the $\Delta^2$ map for the whole field in order to
identify regions with large small-scale inhomogeneities.  The outcome
of this analysis (see Figs.~\ref{fig:15} and \ref{fig:16}) support some
results recently obtained for the Pipe nebula \citep{L08}:
inhomogeneities are mostly present in high column density regions,
while at low extinctions (approximately below $A_K < 0.5 \mbox{ mag}$)
substructures are either on scales large enough to be detected at our
resolution ($3 \mbox{ arcmin}$), or are negligible.  The range of
values spanned by the $\Delta^2$ map is quite impressive: it reaches
$(1.5 \mbox{ mag})^2$ in the $\rho$ Ophiuchi core, indicating an
extraordinary amount of small-scale substructures present there.  The
Lupus region shows less pronounced small-scale inhomogeneities, with a
maximum $\Delta^2 \simeq (0.66 \mbox{ mag})$ found in the Lupus~3 core
(not shown here).  The Lupus $\Delta^2$ map at low galactic latitudes
shows significant amount of substructures, possibly due to the
presence of other molecular clouds (see below).

Figures~\ref{fig:17} and \ref{fig:18} shows the average $\Delta^2$ as
a function of the local extinction $A_K$ for the Ophiuchus and the
Lupus clouds.  As shown by these plot, substructures start to play a
significant role at relatively large column densities and are
negligible at low $A_K$.  In particular, the dashed lines in these
plots, representing the average value of $\Delta^2$ in bins of $0.05
\mbox{ mag}$ in $A_K$, should be compared with the average variance
$\mathrm{Var}\bigl( \hat A^{(n)}_K \bigr)$ on the estimate of $\hat
A_K$ from a single star, which is approximately $0.033 \mbox{ mag}^2$.
From Figs.~\ref{fig:17} and \ref{fig:18} we thus see that local
inhomogeneities start to be the prevalent source of errors in
extinction maps for $A_K > 0.8 \mbox{ mag}$ for the Ophiuchus cloud.
For the Lupus complex, apparently this happens before, but a more
detailed analysis shows that this result is mostly due to the rapid
increase of $\Delta^2$ at low galactic latitudes (which, in turn, is
probably due to other intervening clouds).

Diagrams such as the ones presented in Figs.~\ref{fig:17} and
\ref{fig:18} are invaluable to understand the small-scale properties
of molecular clouds, but clearly they can only be interpreted with a
detailed model.  A thorough analysis of these results goes beyond the
scope of this paper, and in any case is hampered by the relatively
coarse resolution achievable using 2MASS data.  Still these data allow
us to confirm that the observed scatter in column density is due to
small-scale substructure and not to other effects such as foreground
star contamination.  The latter, typically, show up in the
$A_K$-$\Delta^2$ diagram as separate trails with parabolic shape that
divert from the main $\Delta^2 = 0$ locus of points (cf. Fig.~9 in
\citealp{1994ApJ...429..694L}, where however ${\hat\sigma}_{\hat A_K}
\sim \sqrt{\Delta^2}$ is plotted, so that the parabolic shape becomes
a line).  This pattern typically is well discernible, and in absence
of other sources of inhomogeneities produces a clear signature that is
not compatible with the results obtained in Figs.~\ref{fig:17} and
\ref{fig:18}.  In summary, the data seem to indicate a genuine
presence of small-scale inhomogeneities.  Although a specific,
model-dependent analysis would be needed to better understand their
origin, the evidence we have suggests that it is not unlikely that
these substructures are associated to steep gradients (or possibly
unresolved clumps) in the extinction map.

\subsection{Cloud structure functions}
\label{sec:cloud-struct-funct}

\begin{figure}[!tbp]
  \begin{center}
    \includegraphics[width=\hsize, bb=19 7 325 242]{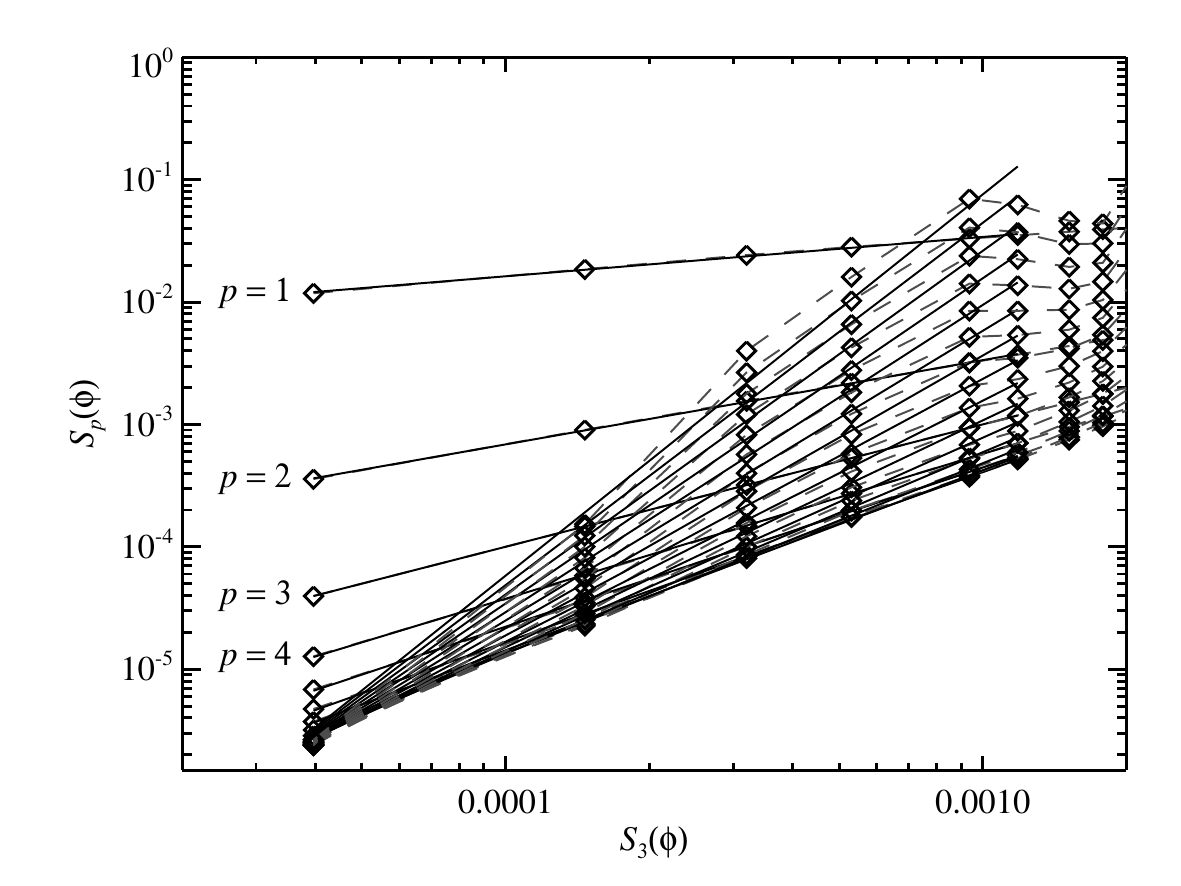}
    \caption{The structure functions for the Ophiuchus complex.  The
      diamonds show the measured $S_p(\phi)$ as a function of
      $S_3(\phi)$, with $p$ increasing from 1 to 20; the dashed lines
      connect diamonds with the same $p$ and different $\phi$;  the
      solid lines are the best exponential fits.  The first four
      values of $p$ are marked at the left of the corresponding fit.}
    \label{fig:19}
  \end{center}
\end{figure}

\begin{figure}[!tbp]
  \begin{center}
    \includegraphics[width=\hsize, bb=0 17 328 406]{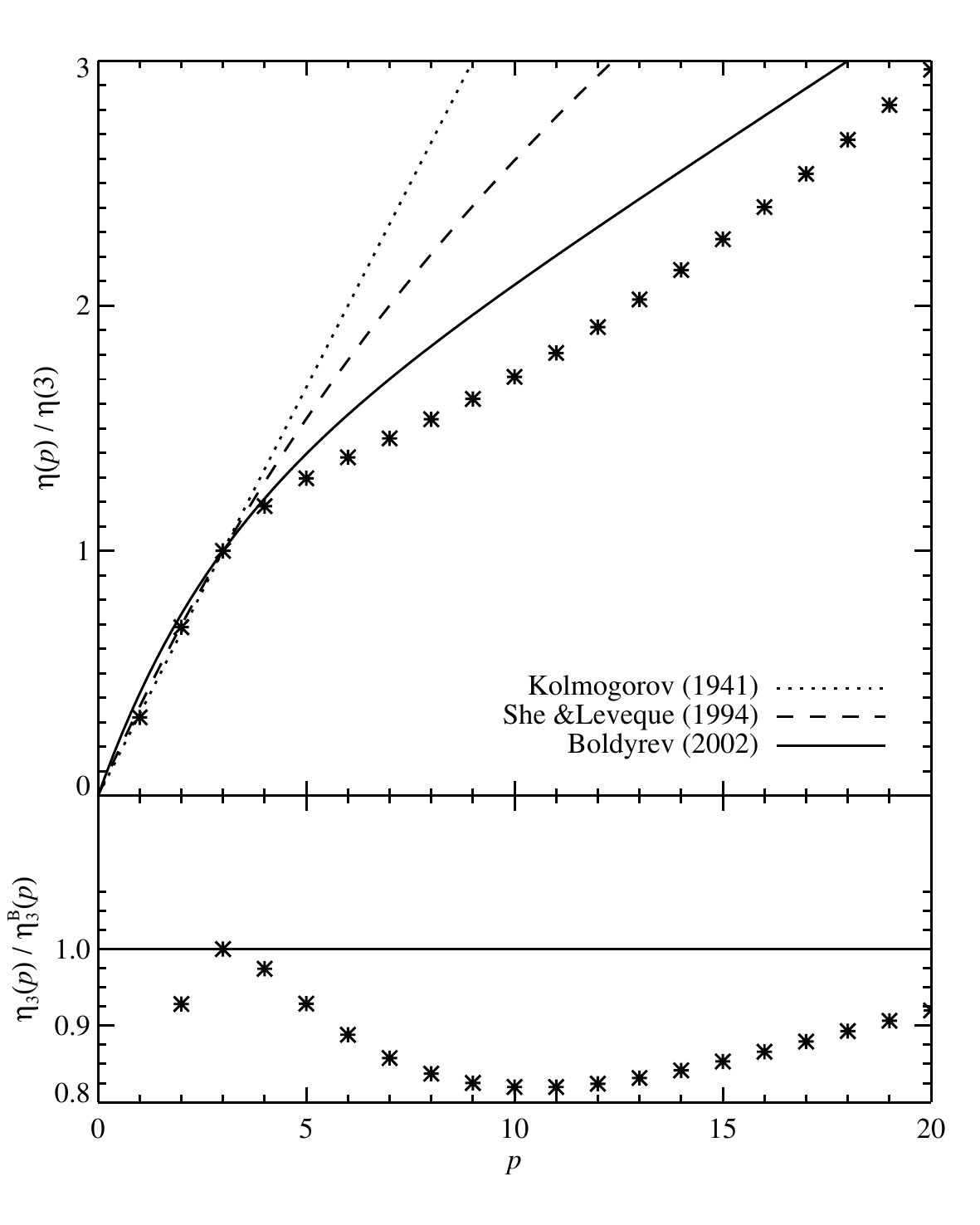}
    \caption{The scaling index ratio $\eta(p) / \eta(3)$ as a function
      of the structure function order $p$ for the Ophiuchus cloud [cf.\
      Eq.~\eqref{eq:26}], with the predictions from three theoretical
      models.  The bottom panel shows the ratio between the observed
      scaling index and the one predicted by Boldyrev (2002).}
    \label{fig:20}
  \end{center}
\end{figure}

\begin{figure}[!tbp]
  \begin{center}
    \includegraphics[width=\hsize, bb=0 17 328 406]{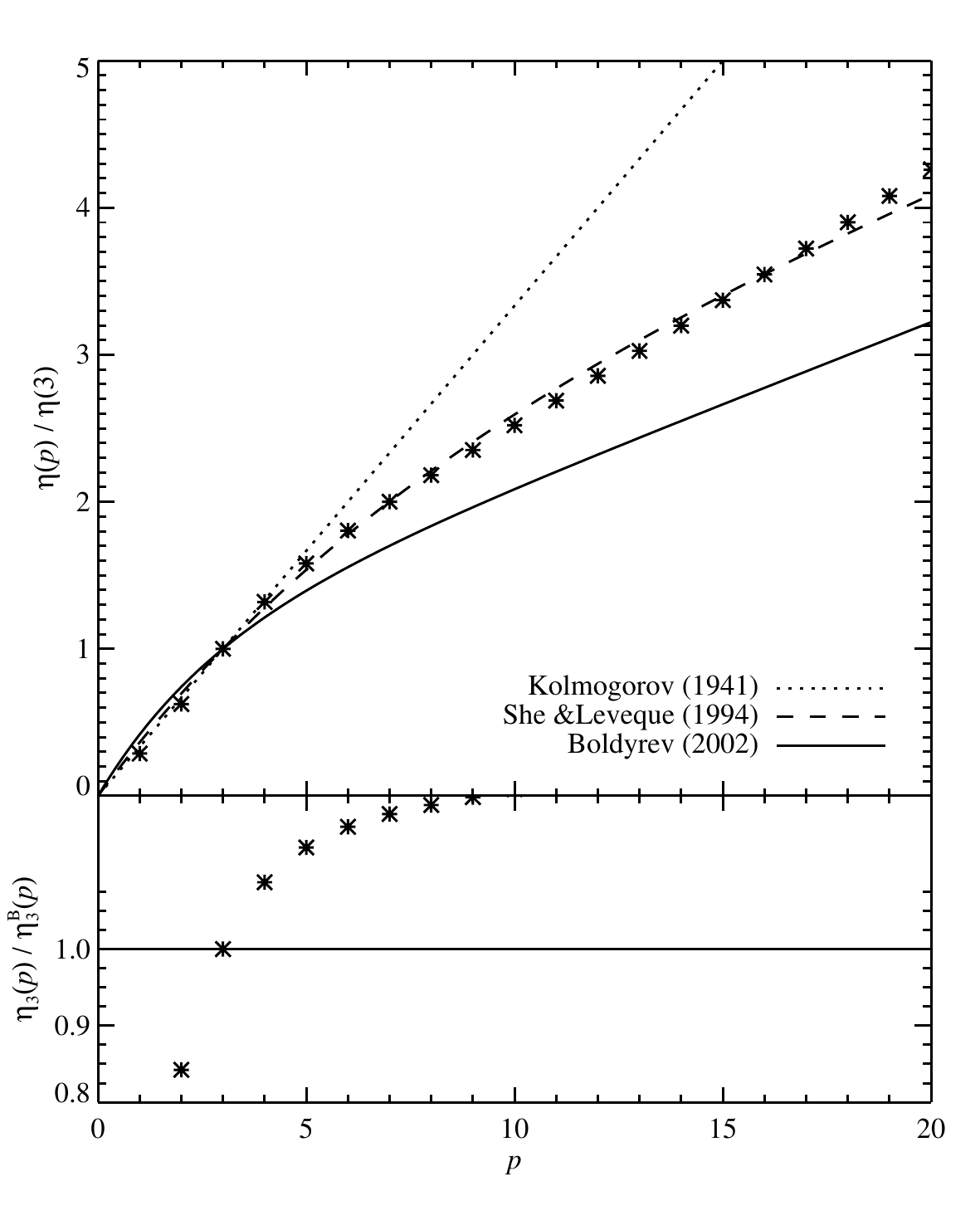}
    \caption{Same as Fig.~\ref{fig:20}, but for the Lupus cloud.  In
      this case the \cite{1994PhRvL..72..336S} model seems to able to
      reproduce very well the data.  Note that no significant
      difference in the scaling index is observed when the low
      galactic latitude regions of the cloud are discarded in the
      analysis.}
    \label{fig:21}
  \end{center}
\end{figure}

The structure functions of the extinction map of a molecular cloud are
defined as
\begin{equation}
  \label{eq:25}
  S_p(\phi) \equiv \bigl\langle \left\lvert A_K(\vec\theta) - A_K(\vec
    \theta + \vec \phi) 
  \right\rvert^p \bigr\rangle \; ,
\end{equation}
where the average is carried over all positions $\vec\theta$ and all
directions for $\vec\phi$ (note that for $p = 2$, the structure
function $S_2(\phi)$ is the usual two-point correlation function of
the extinction map $A_K$).  

Starting from \citet{Kolmogorov}, one of the focuses of turbulence
theory have been the \textit{statistical\/} properties of the
\textit{velocity\/} field.  In a wide range of length scales, known as
``inertial range,'' the effects of both external forces (which are
taken to act on very large scales) and viscosity (which plays a role
on very small scales) are negligible.  As a result, in this range the
energy of large scale flows is merely transferred to smaller scales
until viscous effects become important, a process named ``energy
cascade.''  This naturally leads to random, isotropic motions (because
the imprint of the large scale flows is likely to be lost during the
energy cascade), which can be studied with the help of velocity
structure functions, defined similarly to $S_p(\phi)$ of
Eq.~\eqref{eq:25}.  In his seminal paper, \citet{Kolmogorov}
considered the first two structure velocity functions $S_2$ and $S_3$,
and showed that both are simple power laws of the separation $\phi$,
with exponents $2/3$ and $1$.  Since Kolmogorov's theory implicitly
assumes that turbulence is statistically self-similar at different
scales, one can actually extend this result to any order $p$, and show
that the structure functions of the velocity field must be simple
power laws of the angular distance parameter $\phi$, i.e.
\begin{equation}
  \label{eq:26}
  S_p(\phi) \propto \phi^{\eta(p)} \; ,
\end{equation}
where $\eta(p) = p / 3$ in the simplest turbulent model considered by
Kolmogorov.  However, experiments and numerical simulations have shown
that, although Eq.~\eqref{eq:26} applies to a variety of turbulent
flows at high Reynolds number, $\eta(p)$ substantially deviates from
linearity at higher orders $p$, a phenomenon often referred to as
``intermittency.''  \citet{1994PhRvL..72..336S} have proposed a model
for incompressible turbulence based on intermittency, further extended
by \citet{2002ApJ...569..841B} and tested with numerical simulations
of supersonic turbulence \citep{2002ApJ...573..678B}.

Although no equivalent models are available for the (projected)
density of molecular clouds, we follow \citet{2002ApJ...580L..57P} and
assume that the same scaling law $\eta(p)$ applies to both the
velocity and density fields.  We analysed thus the structure functions
$S_p(\phi)$ and scaling law $\eta(p)$ for both the Ophiuchus and Lupus
complexes.  Figure~\ref{fig:19} shows the observed structure functions
up to $p = 20$ on the Ophiuchus complex, together with the best
exponential fits (which are found to be appropriate in our case).  The
dependence of the exponent $\eta(p)$ of the best fit on the structure
function order $p$ is shown in Fig.~\ref{fig:20}, together with the
predictions of the three turbulent models discussed above.  Note that
in this analysis we considered the ratio $\eta(p) / \eta(3)$, which
according to \citet{1993PhRvE..48...29B} \citep[see
also][]{1994PhRvL..73..959D} should show a universal behaviour also at
relatively small Reynolds numbers.  While the Boldyrev model seems to
fit reasonably well the data for the Taurus molecular cloud
\citep{2002ApJ...580L..57P}, the fit seems to be very poor for the
Ophiuchus and Lupus complexes (Figs.~\ref{fig:20} and \ref{fig:21}).

Despite the fact that the interpretation of these results can be
complicated by many factors, the data analysed here seems to indicate
that the Ophiuchus and Lupus clouds have two intrinsically different
structure functions.  In addition, Figs.~\ref{fig:20} and \ref{fig:21}
show that one of the currently favoured turbulent models, the
\citet{2002ApJ...569..841B} model, can not describe accurately the
large-scale structure of these molecular clouds.  Interestingly, for the
Lupus complex a good description of the scaling index ratio $\eta(p)
/ \eta(3)$ is instead given by the \citet{1994PhRvL..72..336S} model,
with relatively small deviations over the whole range of $p$ values
investigated here.  Note that, by construction, the structure function
index $\eta(p)$ is left unchanged by a simple linear (affine)
transformation $A_K(\vec \theta) \mapsto \alpha A_K(\vec \theta) +
\beta$ of the extinction map, and thus is insensitive to errors on the
zero-point of the extinction (control field) and on the reddening law.
Similarly, the structure function index is independent of the distance
of the cloud: in other words, if two physically identical cloud
located at different distances $d$ and $d'$ will have structure
functions that differ only by a scaling factor:
\begin{equation}
  \label{eq:27}
  S'_p(\phi) = S_p(k \phi) \; ,
\end{equation}
where $k = d / d'$, we have
\begin{equation}
  \label{eq:28}
  \phi^{\eta'(p)} \propto S'_p(\phi) = S_p(k \phi) \propto 
  k^{\eta(p)} \phi^{\eta(p)} \propto \phi^{\eta(p)} \; ,
\end{equation}
or $\eta'(p) = \eta(p)$.  Note, however, that if two clouds located at
different distances and with the same structure function index
$\eta(p)$ (or even with the same structure functions $S_p$) are
analysed jointly in the same field, the deduced structure functions
will in general exhibit a non-trivial behaviour, and will not be
simply described as power laws as in Eq.~\eqref{eq:26}.  Since this
paper focuses on two wide molecular cloud complexes, this point might
potentially affect the results and partially be responsible for the
poor fits displayed in Figs.~\ref{fig:20} and \ref{fig:21}.  Hence, we
decided to repeat the analysis in the core of Ophiuchus (see window
marked with solid line in Fig.~\ref{fig:9}) which is most likely
composed by a single cloud at a well determined distance
\citep[e.g.][]{2008A&A...480..785L}, but the results obtained are
completely consistent to the ones shown in Fig.~\ref{fig:20}.  This,
indirectly, confirms that for the purposes of the calculation of the
cloud structure function, there is no significant difference in using
the whole Ophiuchus complex or a smaller subset centered on its core,
or equivalently that most likely the poor fit observed for this cloud
with the turbulent models considered here is not due to the
overlapping, distinct clouds located at different distances.  Finally,
we stress that the conclusions reported here seems to be robust
against cuts of the star catalogue (see
Sect.~\ref{sec:nicer-absorpt-map}) and to changes in the resolution of
our maps.

\section{Mass estimate}
\label{sec:mass-estimate}

\begin{table}
  % Oph  25386    8330
  % Lup  36489   18198
  % P1    9247    7568
  % P2   14821   14254
  \centering
  \begin{tabular}{lcc}
    Cloud            & Total mass & Cloud mass \\
    \hline
    Ophiuchus        & $25\,400 \mbox{ M}_\odot$ & $\phantom{0}8\,300 \mbox{ M}_\odot$ \\
    Lupus            & $36\,500 \mbox{ M}_\odot$ & $18\,200 \mbox{ M}_\odot$ \\
    Pipe             & $14\,800 \mbox{ M}_\odot$ & $14\,200 \mbox{ M}_\odot$ \\
    Pipe (corrected) & $\phantom{0}9\,200 \mbox{ M}_\odot$ & $\phantom{0}7\,600 \mbox{ M}_\odot$ \\
  \end{tabular}
  \caption{The masses of the Ophiuchus, Lupus, and Pipe dark
    complexes.  For the Pipe, the ``corrected'' values reported in the
    last line refer to the extinction map corrected for the $0.15
    \mbox{ mag}$ plateau. The two columns refers to the whole fields
    considered here, and to the regions with $A_K > 0.2 \mbox{ mag}$ only.}
  \label{tab:2}
\end{table}

The cloud mass $M$ can be derived from the $A_K$ extinction map using the
following simple relation
\begin{equation}
  \label{eq:29}
  M = d^2 \mu \beta_K \int_\Omega A_K(\vec\theta) \, \diff^2 \theta \; ,
\end{equation}
where $d$ is the cloud distance, $\mu$ is the mean molecular weight
corrected for the helium abundance, $\beta_K \simeq 1.67 \times
10^{22} \mbox{ cm}^{-2} \mbox{ mag}^{-1}$ is the ratio $\bigl[
N(\mathrm{H\textsc{i}}) + N(\mathrm{H}_2) \bigr] / A_K$
(\citealp{1979ARA&A..17...73S}; see also \citealp{1955ApJ...121..559L,
  1978ApJ...224..132B}), and the integral is evaluated over the whole
field $\Omega$.  Assuming a standard cloud composition ($63\%$
hydrogen, $36\%$ helium, and $1\%$ dust), we find $\mu = 1.37$ and
total masses $M = (25 \, 400 \pm 2 \, 500) \mbox{ M}_\odot$ for
Ophiuchus, and $M = (36 \, 500 \pm 3 \, 600) \mbox{ M}_\odot$ for
Lupus (cf.\ Table~\ref{tab:2}).  The error in both cases is mainly due
to the uncertainty on the distance of the cloud (for both clouds we
used our new distance measurements, see
\citealp{2008A&A...480..785L}); in addition, for Lupus there is an
additional uncertainty (not included in our error budget) due to the
possible projection of other clouds on the line of sight.  We also
considered the total mass of the Pipe nebula from our data.  As shown
by Fig.~\ref{fig:6}, the Pipe nebula is located in an area of fairly
large extinction, and as a result it is reasonable to subtract from
the extinction map of the Pipe a constant value, representing the
``plateau'' where the Pipe nebula is located.  Clearly, an exact,
physically meaningful definition of the value cannot be provided from
the available data, because there is no simple way to disentangle the
effects of different cloud structures located at different distances
from near-infrared extinction measurements alone.  Hence, we decided
here to derive the value of the ``plateau'' extinction from the same
area used as a control field in Paper~II.  This choice is reasonable,
because it selects the lowest extinction values on the area around the
Pipe nebula, and in addition produces results that are comparable to
the ones provided in Paper~II.  Finally, additionally we considered
only regions above $0.2 \mbox{ mag}$ of $K$-band extinction: this
choice allows us to concentrate on the real structures present in the
cloud complexes and to avoid the diffuse, low-density material
surrounding them.  The results obtained for the various combinations
discussed here are reported in Table~\ref{tab:2}.

\begin{figure}[!tbp]
  \begin{center}
    \includegraphics[width=\hsize, bb=5 7 331 239]{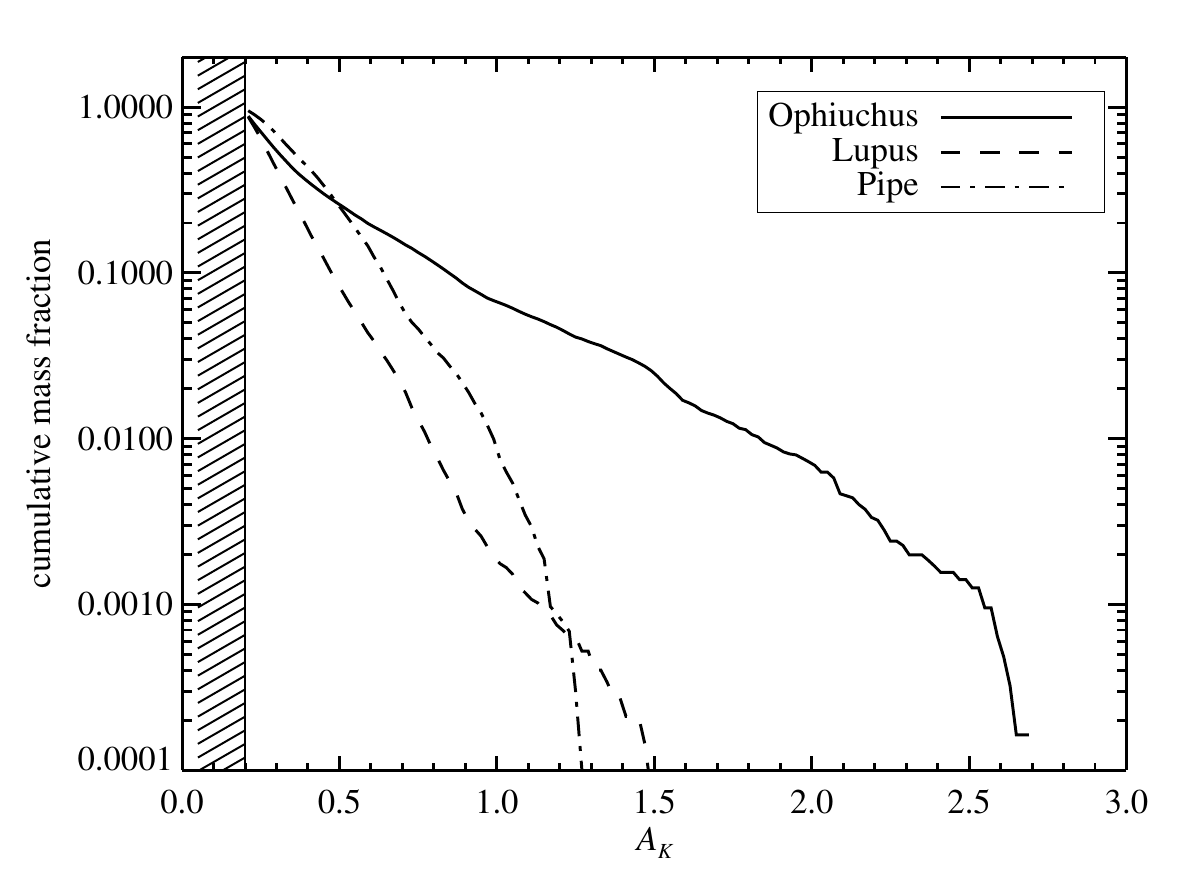}
    \caption{The cumulative mass enclosed in isoextinction contours
      for the Ophiuchus, Lupus, and Pipe molecular clouds.  All plots
      have been obtained from the extinction map shown in
      Fig.~\ref{fig:6}, and have thus the same resolution limit.  Note
      that the difference between the Pipe nebula cumulative mass
      function shown here and the one reported in Fig.~27 of Paper~II
      is essentially due to the different resolution of the two
      extinction maps.}
    \label{fig:22}
  \end{center}
\end{figure}

Figure~\ref{fig:22} shows the relationship between the integrated mass
distribution and the extinction in $A_K$; for the Pipe nebula, we used
the ``corrected'' extinction with the plateau subtracted.  Note that
regions with extinction larger than $A_K > 1 \mbox{ mag}$ account for
less than $1\%$ of the total mass in the whole field.  Hence, we do
not expect any significant underestimation in the cloud mass due to
unresolved dense cores.

More significantly Fig.~\ref{fig:22} also shows a clear difference
between the relative frequencies of extinction characterizing the
Ophiuchus, Lupus and Pipe clouds.  Ophiuchus has a considerably higher
fraction of dust at high extinction ($A_K > 1.0 \mbox{ mag}$) than
either Lupus or the Pipe.  This reinforces the result of
Sect.~\ref{sec:cloud-struct-funct} that shows a significant difference
in the structural properties of the Ophiuchus and Lupus clouds at high
order $p$ of the structure function.  The scaling index curve for
Ophiuchus (Fig.~\ref{fig:20}) is significantly flatter than the
corresponding curve for the Lupus cloud (Fig.~\ref{fig:21}) and the
predictions of turbulence theory.  Inspection of Eq.~\eqref{eq:25}
indicates that this difference is a direct result of the Ophiuchus
cloud having a larger fraction of its material at high extinction
compared both to that in the Lupus cloud or that predicted by standard
turbulence theory.

It is interesting to note in this context that it has been known for
some time \citep[e.g.][]{1992ApJ...393L..25L} that star formation
occurs almost exclusively in dense ($n(\mbox{H}_2) > 10^4 \mbox{
  cm}^{-3}$), high extinction ($A_V > 10 \mbox{ mag}$) gas.  Thus, it
is not surprising that the Ophiuchus cloud is a very active star
formation complex, the site of a relatively rich embedded cluster.  In
contrast, Lupus is characterized by more modest levels of star
formation \citep[e.g.][]{2005ApJ...629..276T} and the Pipe cloud is
noted for its nearly complete absence of star forming activity
\cite[e.g.][]{BlueBook}. The closer correspondence of the Lupus
structure function to the predictions of turbulence theory
(Fig.~\ref{fig:21}) compared to that of Ophiuchus (Fig.~\ref{fig:20})
may indicate that standard turbulence inhibits the large scale
production of dense gas and suppresses active star and cluster
formation.

\section{Conclusions}
\label{sec:conclusions}

The main results of this paper can be summarized as follows:
\begin{itemize}
\item We used approximately $42$ million stars from the 2MASS point
  source catalog to construct a $1\,672$ square degrees \textsc{Nicer}
  extinction map of the Ophiuchus and Lupus dark nebul\ae.  The map has
  a resolution of $3\mbox{ arcmin}$ and an average $2\sigma$ detection
  level of $0.5$ visual magnitudes.
\item We considered in detail the effect of sub-pixel inhomogeneities,
  and derived an estimator useful to quantify them.  We also showed
  that inhomogeneities play a significant role only in the densest
  cores with $A_K > 6$--$8 \mbox{ mag}$.
\item We derived the structure functions of both dark clouds and
  compared them with several theoretical models.  We could not find any
  reasonable fit of the Ophiuchus data with models, while the Lupus
  scaling index ratio is well described in terms of the
  \citet{1994PhRvL..72..336S} turbulent model.
\end{itemize}

\acknowledgements 

We thank the anonymous referee for many useful comments and
suggestions.  This research has made use of the 2MASS archive,
provided by NASA/IPAC Infrared Science Archive, which is operated by
the Jet Propulsion Laboratory, California Institute of Technology,
under contract with the National Aeronautics and Space Administration.
CJL acknowledges support from NASA ORIGINS Grant NAG 5-13041.

\bibliographystyle{aa} 
\bibliography{../dark-refs}

\end{document}